\documentclass[journal]{IEEEtran}
\usepackage{graphicx}

\ifCLASSINFOpdf
\else
\fi
\usepackage{amsmath}
\usepackage{array}
\usepackage{cite}  %% ????
\usepackage{amsmath,amssymb,amsfonts}
\usepackage{algorithmic}
\usepackage{graphicx}
\usepackage{textcomp}
\usepackage{color}

\usepackage{tabularx}
\usepackage{booktabs}
\usepackage{comment}

\newlength\myindent
\newcommand\bindent{%
  \begingroup
  \setlength{\itemindent}{\myindent}
  \addtolength{\algorithmicindent}{\myindent}
}
\newcommand\eindent{\endgroup}
\usepackage{float}
\usepackage{placeins}
\usepackage{caption,setspace}
\usepackage{subcaption}

\begin{document}

\title{An algorithm for DC segmentation of AC power systems to mitigate electromechanical oscillations}

\author{Mathieu~Robin,
        Javier~Renedo (Senior Member, IEEE),         
        Juan Carlos~Gonzalez-Torres,
        Aurelio~Garcia-Cerrada (Senior Member, IEEE),  
        Abdelkrim~Benchaib,  
        and~Pablo~Garcia-Gonzalez  
        % <-this % stops a space
\thanks{This work is supported by the French Government under the program Investissements d’Avenir (ANEITE-002-01).}% <-this % stops a space
}

\markboth{Preprint}%
{Shell \MakeLowercase{\textit{et al.}}: Bare Demo of IEEEtran.cls for IEEE Journals}

\maketitle

\begin{abstract}\label{sec:asbtract}
In the last decades, various events have shown that electromechanical oscillations are a major concern for large interconnected Alternative Current (AC) power systems. DC segmentation - a method that consist in turning large AC grids into a set of asynchronous AC clusters linked by Direct Current (DC) links - is a promising solution to mitigate this and other issues. However, no systematic segmentation procedure for a given AC power system exists so far. This paper aims at filling this gap and proposes an algorithm for DC segmentation for a given AC power system to mitigate electromechanical oscillations. In this proposal, DC segmentation is implemented with High Voltage Direct Current links based on Voltage Source Converters (VSC-HVDC). The algorithm uses small-signal stability techniques and the concept of dominant inter-area oscillation paths to stop the main inter-area mode of the power system. The algorithm will be explained using a six-generator  test system and it will then be used on the Nordic 44 test system. The proposed algorithm for DC segmentation has been validated by means of non-linear time-domain simulation and small-signal stability analysis (SSSA).
\end{abstract}


% \fi
\IEEEpeerreviewmaketitle

\section{Introduction}\label{sec:Intro}
\IEEEPARstart{A}lternating Current (AC) technology is currently the dominant technology for the transmission and distribution of electricity. However, the transmission capability with High Voltage Alternating Current (HVAC) is limited by some technical aspects that can be overcome by the High Voltage Direct Current (HVDC) transmissiontechnology that has been developed in the last decades \cite{vanhertemMultiterminalVSCHVDC2010,hertemHVDCGridsOffshore2016,luscanVisionHVDCKey2021}. Thus, many point-to-point HVDC links are already in operation around the world, and even more are planned or under construction \cite{entso-eGridMap2021,alassiHVDCTransmissionTechnology2019}. This means that the electrical grids are evolving toward a hybrid HVAC/HVDC power system with a growing share of HVDC transmission. 

In parallel with the development of HVDC links, power systems have also been more and more interconnected creating large synchronous electrical grids. This action sums the interconnected systems' inertia values, improves frequency stability and increases the reliability of the resulting power systems. However, with the introduction of new economical objectives, power systems are operated closer to their stability limits and the risk of severe disruptions is increasing rapidly. For example, in the last decade, various events have shown that electromechanical oscillations are now a bigger threat than ever \cite{entso-eAnalysisCEInterArea2017}.

In situations where electromechanical oscillations become a threat to the stability of the system, corrective measures need to be taken. For example, intentional islanding will disconnect selected lines to split a grid into a set of stable areas – called islands – that are reconnected once the fault and its propagation have been properly addressed. The key issue of this technique is the selection of the islands in order to actually stop the propagation of the fault and to limit the time needed for the reconnection. Various approaches have been proposed including graph partitioning-based method \cite{liControlledPartitioningPower2010} slow coherency-based method \cite{youSlowCoherencyBasedIslanding2004} and ordered binary decision diagram (OBDD)-based approach \cite{kaisunSplittingStrategiesIslanding2003}.

However, intentional islanding is only an operational action. Some planning actions can also be taken to limit the risk of stability issues, for example the construction of embedded HVDC links which, thanks to their high controllability, can contribute to stabilise a power system. In particular, many papers propose various control strategies of VSC-HVDC to damp electromechanical oscillations \cite{elizondoInterareaOscillationDamping2018} or to improve transient stability \cite{gonzalez-torresTransientStabilityPower2020}.

Another promising application of VSC-HVDC to increase the stability of a power system is the DC segmentation. This concept was first proposed in \cite{clarkApplicationSegmentationGrid2008} with the following definition: DC-\emph{Segmentation involves breaking large grids […] into smaller ac sectors interconnected by BtB (Back-to-Back HVDC links) and HVDC transmission lines}. DC segmentation and intentional islanding have some similarities; indeed, they both consist in splitting a big electrical grid into a set of smaller asynchronous clusters. Yet, they are clearly different: intentional islanding is a corrective measure used as a last resort to stop the propagation of a fault and lasts only a limited period of time. While DC-segmentation is a planning action that creates permanent clusters to improve the stability of the overall system. Additionally, in the case of DC-segmentation, clusters are linked together by DC links while it is usually not the case between the clusters created by intentional islanding. 

DC-segmentation has the capacity to limit the propagation of perturbations \cite{clarkSofteningBlowDisturbance2008, fangBTBDCLink2009,shamiAnalysisDifferentPower2015}. Thus, it could limit the risk of cascading failures and black-outs \cite{clarkApplicationSegmentationGrid2008}. This effect  has been confirmed in \cite{mousaviAssessmentHVDCGrid2013} with DC links controlled with frequency support and in \cite{gomilaAnalysisBlackoutRisk2023} with DC links modelled as AC lines with controllable impedance. Both papers show that DC-segmentation greatly reduce the risk of large scale blackouts.

The work in \cite{robinDCSegmentationPromising2021} presented a comprehensive analysis of the impact of DC segmentation on transient stability, electromechanical-oscillation damping and frequency stability in power systems. It showed that DC segmentation could improve transient stability and the damping of electromechanical oscillations (the DC segment acts as a firewall), while frequency stability is deteriorated (because each of the resulting asynchronous AC clusters has less inertia and frequency control deteriorates).

ENTSO-E has also identified DC-segmentation as a potential solution to improve power system stability \cite{entso-eHVDCLinksSystem2019}. A first DC segmentation project was carried out in China in 2016~ \cite{fairleyWhySouthernChina2016} to prevent potential overloading of the inter-area AC lines in case of contingency.

Since the advantages of DC segmentation have been accessed, the problem of how to select the DC segmentation scheme for a given power system rises. The only previous work in this mater are the patent in \cite{el-gasseirElectricityMarketorientedDcSegmentation2012} that proposes to follow the market boundaries without consideration 
of power system stability and \cite{chengOptimalDCSegmentationMultiInfeed2016} that compares different segmentation schemes to limit the commutation failure of the infeed line commutated converter of the system but without justifying the
initial selection of the segmentation candidates. Thus, no previous work has addressed the problem of how to select the optimal DC segmentation scheme for a given power system based on technical aspects. 

Therefore, the aim of this paper is to fill this gap and it proposes an algorithm that, for a given power system, gives a DC-segmentation architecture to improve electromechanical-oscillation damping. The algorithm relies on separating the groups of generators oscillating against each other using the mode shapes of the system. The exact boundaries of the clusters are then identified using the concept of dominant inter-area oscillation path, which was proposed in \cite{chompoobutrgoolIdentificationPowerSystem2013}. The algorithm selects the AC lines to be replaced by VSC-HVDC links, leading to a DC-segmented system with asynchronous AC clusters. 

In the methodology presented in \cite{chompoobutrgoolIdentificationPowerSystem2013}, the main characteristics of the inter-area oscillation paths were identified, and a process to find this path has been defined. However, the process is not fully automatised and requires human interaction. In this paper, and as part of the proposed algorithm for DC-segmentation, a method to identify the inter-area oscillation path has been developed, by adding some specific features to the guidelines presented in \cite{chompoobutrgoolIdentificationPowerSystem2013}, in order to fully determine the path without human intervention. Hence, the method of obtaining the inter-area oscillation path is also a contribution of this paper and could be used, not only for DC segmentation, but also for other applications.

The rest of the paper is organised as follows. Section \ref{sec:Inter-area_oscillation_path} describes the concept of inter-area oscillation paths and introduces its application to DC segmentation targeted to damp electromechanical oscillations. It provides the theoretical bases for the proposed algorithm. Section~\ref{sec:DC_segmentation_algorithm} introduces the proposed algorithm for DC segmentation. Section~\ref{sec:results} presents the results when the proposed algorithm is applied to the Nordic 44 test system. Section \ref{sec:conclusions} presents the conclusions obtained in this work. Finally, data of the tests systems and VSC-HVDC systems used in this paper are presented in Appendix.

\section{Dominant inter-area oscillation paths and DC segmentation}\label{sec:Inter-area_oscillation_path}
This section presents the fundamentals of inter-area oscillation paths in power systems, based on the work of \cite{chompoobutrgoolIdentificationPowerSystem2013}, and it illustrates heuristically the impact of DC segmentation on inter-area oscillation damping. 

\subsection{Theoretical background}\label{sec:Inter-area_oscillation_path_theor}
It is well known that the electromechanical behaviour of power system can be analysed by a set of non-linear algebraic-differential equations (also known as electromechanical-type or Root-Mean-Square (RMS)-type models)~\cite{kundurPowerSystemStability1993}. The dynamic behaviour of a power system when subject to small disturbances can be analysed using a linearised model of the system around the steady-state operating points \cite{kundurPowerSystemStability1993}. For example, the free response of the system can be written as:
\begin{eqnarray}
  	\boldsymbol{\Delta \dot{x}} & = & \boldsymbol{A \Delta x}, \;\;\boldsymbol{\Delta x}=[\boldsymbol{\Delta \delta}, \mbox{\space} \boldsymbol{\Delta \omega}, \mbox{\space} \boldsymbol{\Delta z}]^{T} \label{eq:SSA_lin_model1}\\
   \boldsymbol{\Delta y} & = & \boldsymbol{C \Delta x} \label{eq:SSA_lin_model2}
\end{eqnarray}
where $\boldsymbol{\Delta x} \in \mathbb{R}^{n_x \times 1}$ is the state vector (increments with respect to the operating point), $\boldsymbol{\Delta y}$ is the output vector and $\boldsymbol{A} \in \mathbb{R}^{n_x \times n_x}$ is the state matrix of the system. The state vector contains, explicitly, the rotor angles ($\boldsymbol{\Delta \delta}$) and speeds of the generators ($\boldsymbol{\Delta \omega}$), and the rest of state variables ($\boldsymbol{\Delta z}$), in order to analyse electromechanical oscillations\cite{chompoobutrgoolIdentificationPowerSystem2013}. 

In addition, if $\lambda_k$ is an eigenvalue of matrix $\boldsymbol{A}$ and $\boldsymbol{v_k}$ is its associated righ eigenvector (i.e. $\lambda_{k} \boldsymbol{v_{k}}= \boldsymbol{A v_{k}}$), \cite{kundurPowerSystemStability1993} shows that: 
\begin{equation}
\boldsymbol{ \Delta x}= \sum_{\forall k} \left( \boldsymbol{v_k} z_k(0) e^{\lambda_k t} \right)
\end{equation}
where each $z_k(0)$ is a linear combination of the initial conditions of the state variables. 

Therefore, right eigenvector $\boldsymbol{v_k}$ ``shape'' the way in which a system mode (eigenvalue) affects the time response of each of the state variables. Specifically, if the elements of vector $\boldsymbol{v_k}$ are complex numbers, the phases of those numbers affect the relative phases of the oscillatory response of the state variables due to $\lambda_k$ and its complex conjugate. This is why right eigenvalues are also known as ``mode shapes''. If two elements of mode shape $\boldsymbol{v_k}$ (e.g. $\boldsymbol{v_k}(i)$ and $\boldsymbol{v_k}(j)$) have similar phases, these two variables are said to be oscillating together while if their phases differ in almost ${\rm 180}^{{\rm o}}$, they are said to oscillate against each other. Using this property, mode shapes of the modes associated to generator speeds have been remarkably useful to analyse generator speed oscillations in power systems~\cite{kleinFundamentalStudyInterarea1991,kundurPowerSystemStability1993,roucoMultiareaAnalysisSmall1993}. Modes related to generator speeds can be identified looking at participation factors~\cite{perez-arriagaSelectiveModalAnalysis1982,vergheseSelectiveModalAnalysis1982,pagolaSensitivitiesResiduesParticipations1989}. 

Matrix $\boldsymbol{C}$ in~(\ref{eq:SSA_lin_model1}) can be further detailed to highlight the way in which state variables affect the time response of the output variables. For example, if bus voltages, bus frequencies and brunch current flows are selected as outputs, one can write:   

\begin{equation}\label{eq:Network_sensitivities2}
     \begin{bmatrix}
    \boldsymbol{\Delta V} \\
    \boldsymbol{\Delta f} \\
    \boldsymbol{\Delta I_f} \\
\end{bmatrix} 
=
\boldsymbol{C}\boldsymbol{\Delta x}
=
  \begin{bmatrix}
\boldsymbol{C_{V}}\\
\boldsymbol{C_{f}}\\
\boldsymbol{C_{I_f}}\\
\end{bmatrix} 
     \begin{bmatrix}
    \boldsymbol{\Delta \delta} \\
    \boldsymbol{\Delta \omega} \\
    \boldsymbol{\Delta z}
\end{bmatrix} 
\end{equation}
where
\begin{eqnarray}\label{eq:Network_sensitivities_C}
\boldsymbol{C_{V}}
 &=&
  \begin{bmatrix}
\boldsymbol{C_{V\delta}} & \boldsymbol{C_{V\omega}} & \boldsymbol{C_{Vz}}\\
\end{bmatrix} 
\\ \nonumber
\boldsymbol{C_{f}}
 &=&
  \begin{bmatrix}
\boldsymbol{C_{f\delta}} & \boldsymbol{C_{f\omega}} & \boldsymbol{C_{fz}}\\
\end{bmatrix} 
\\ \nonumber
\boldsymbol{C_{I_f}}
 &=&
  \begin{bmatrix}
\boldsymbol{C_{I_f \delta}} & \boldsymbol{C_{I_f \omega}} & \boldsymbol{C_{I_f z}}\\
\end{bmatrix} 
\end{eqnarray}
and $\boldsymbol{C_{V}}$,  $\boldsymbol{C}_f$, $\boldsymbol{C_{I_f}}$ are defined in~\cite{vanfrettiAnalysisPowerSystem2010} as Network sensitivity matrices of bus voltages, bus frequencies and branch current flows, respectively.

Given a set of system outputs, observability factors of a system mode $\lambda_k$ are defined as the product of the output-variables network sensitivity matrix by the right eigenvector associated to $\lambda_k$~\cite{vanfrettiAnalysisPowerSystem2010,chompoobutrgoolIdentificationPowerSystem2013}. For example, given the output partition in~(\ref{eq:Network_sensitivities2}), the corresponding observability factors would be:
\begin{equation}\label{eq:obs_vector2}
    \boldsymbol{\phi_{V,k}} = \boldsymbol{C_{V} v_{k}}, \;\; \boldsymbol{\phi_{f,k}} = \boldsymbol{C_{f} v_{k}}, \;\;  \boldsymbol{\phi_{I_f,k}} = \boldsymbol{C_{I_f} v_{k}}
\end{equation}
%where $\boldsymbol{\phi_{V,k}},\boldsymbol{\phi_{f,k}} \in \mathbb{C}^{n \times 1}$, $\phi_{V,k},\phi_{f,k} \in \mathbb{C}^{n \times 1}$

Very much like with mode shapes, the phases of two complex elements of an observability factor will tell which variables oscillate together and which ones oscillate against each other when a system mode is excited. 

The work in \cite{vanfrettiAnalysisPowerSystem2010} uses the term \emph{network mode shape} to refer to vectors in (\ref{eq:obs_vector2}), although the more popular term observability factor will be used in this work, as in~\cite{kundurPowerSystemStability1993}. 

\subsection{Dominant inter-area oscillation  path} \label{sec:path_theory}
\begin{figure}[htb!]
\vspace{-0.3cm}
\centering
\includegraphics[width=0.45\textwidth]{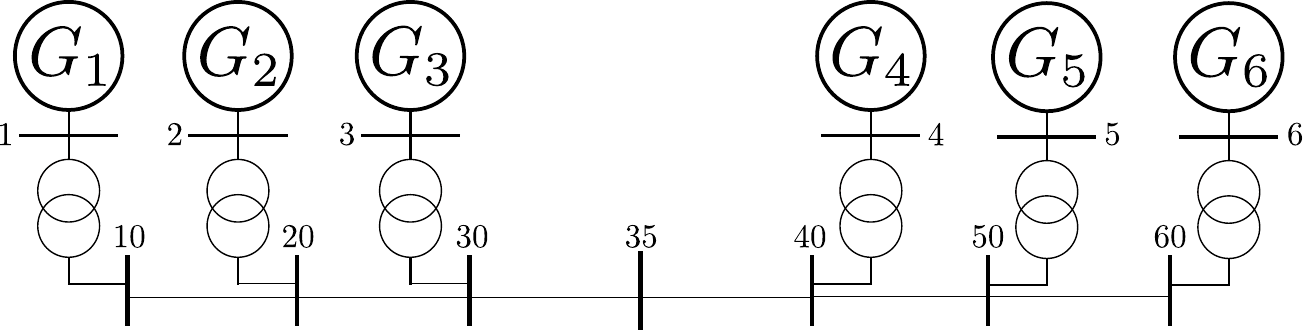}
\caption {Test system 1: Conceptual 6-generator system.}
\label{fig:6g}
\end{figure}

The work in \cite{chompoobutrgoolIdentificationPowerSystem2013} proposed the concept of dominant inter-area oscillation path which can be illustrated using the two-area 6-generator system shown in Figure \ref{fig:6g} (test system 1) with the data reported in Section of \ref{sec:Appendix-6g} the Appendix. 

\begin{figure}[htb!]
\vspace{-0.3cm}
\centering
\includegraphics[width=0.3\textwidth,trim={0.7cm 0.3cm 0.7cm 0.3cm},clip]{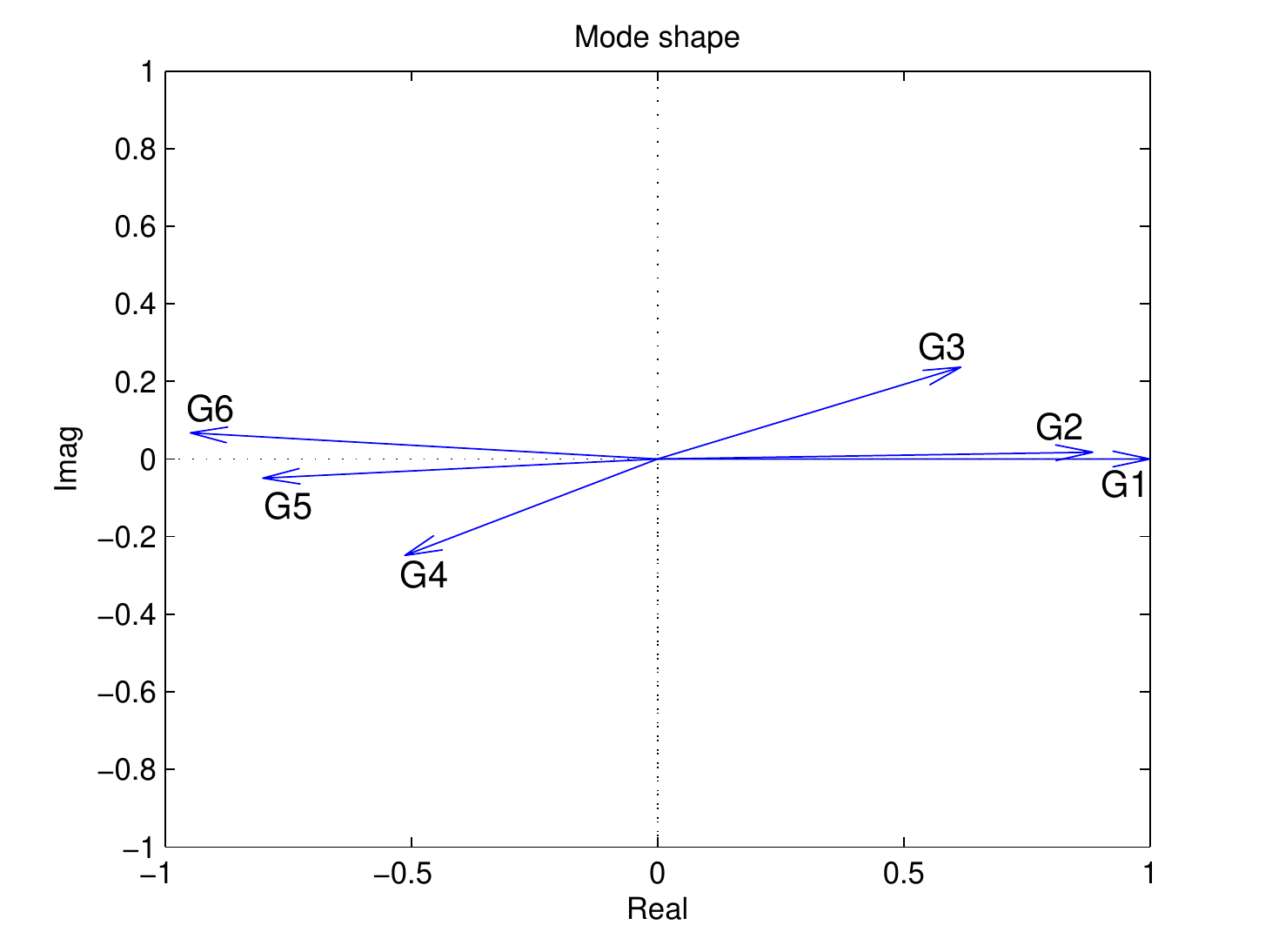}
\caption{Test system 1: Graphical representation of the mode shape of inter-area mode 1.}
\label{fig:MS_6g}
\end{figure}

The small-signal analysis of the system in Fig.~\ref{fig:6g} reveals that it  has an inter-area mode with a damping coefficient of 5.5\% and a frequency of $0.83$ Hz (all system modes are included in Table~\ref{tab:eigv6g}). The generator-speed mode shapes of this mode have been drawn in Fig.~\ref{fig:MS_6g}  which shows that generators G1, G2 and G3 oscillate against generators G4, G5 and G6. 
Fig.~\ref{fig:MS_6g} also shows that the generators further away from the centre of the system (G1 and G6)  are subject to the largest oscillations (the moduli of their mode shapes are the two largest ones) and they are called ``edges'' of the inter-area mode. This mode also affects G2, G3, G4 and G5 whose mode shapes are placed between the two extremes G1 and G6. The buses between 1 and 6 are the dominant path of the inter-area mode 1. 

%%%%%%%%%%%%%%%%%%%%%%%%%%%%%%%%%%%%%%%%%%%%%%%%%%%%%%%%%%%%%%%%%%%%%%%%%%%%%%%%%%%%%%%%%
\begin{table}[htb!]
%\resizebox{\columnwidth}{!}
\centering
\caption{\label{tab:eigv6g} Test system 1: Electromechanical modes.}
\scalebox{0.65}{
\begin{tabular}{@{}cccccc@{}}
\toprule
Mode & Real    & Imag   & Damp. (\%) & Freq. (Hz) & Oscillation \\ \midrule
1	& -0.29	& 5.23	& 5.5	& 0.83	& G1,G2,G3//G4,G5,G6  \\	
2	& -1.17	& 6.35	& 18.1	& 1.03	& G4,G3//G6,G1        \\	
3	& -1.22	& 6.48	& 18.5	& 1.05	& G3,G6//G1,G4,G5     \\	
4	& -1.25	& 6.60	& 18.6	& 1.07	& G2//G1              \\	
5	& -1.27	& 6.51	& 19.2	& 1.06	& G5//G4,G6           \\	  \bottomrule
\end{tabular}%
}
\end{table}
%%%%%%%%%%%%%%%%%%%%%%%%%%%%%%%%%%%%%%%%%%%%%%%%%%%%%%%%%%%%%%%%%%%%%%%%%%%%%%%%%%%%%%%%%%%

\begin{figure} [htb!]
\vspace{-0.1cm}
     \centering
     \begin{subfigure}[b]{0.45\textwidth}
         \centering
         \includegraphics[width=\textwidth,trim={0 0.4cm 0 0.3cm},clip]{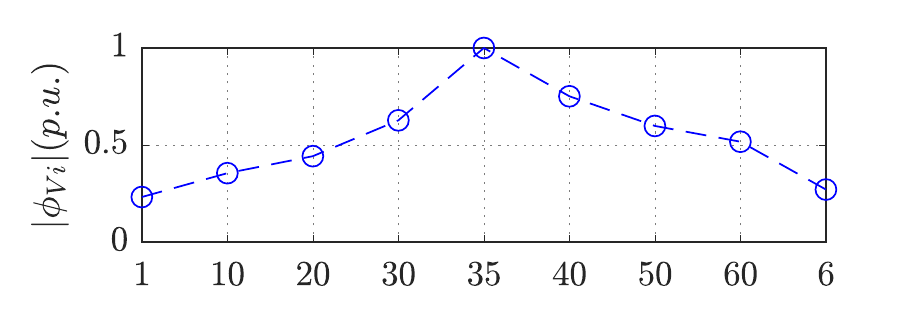}
         \caption{}
         \label{fig:pathV6g_mag}
     \end{subfigure}
     \hfill
     \begin{subfigure}[b]{0.45\textwidth}
         \centering
         \includegraphics[width=\textwidth,trim={0 0.4cm 0 0.25cm},clip]{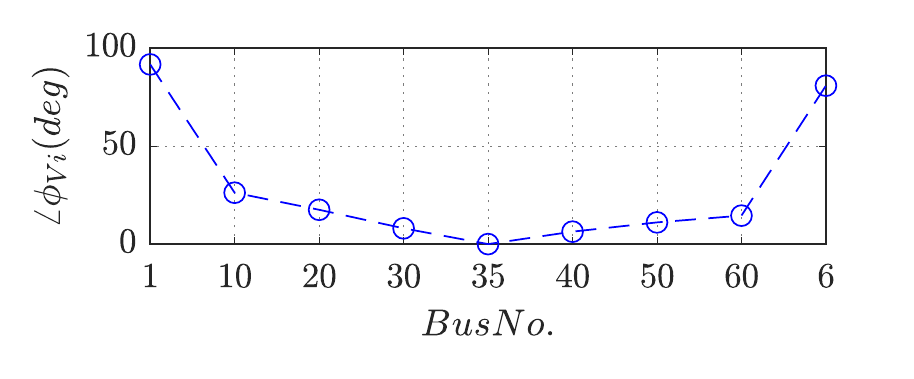}
         \caption{}
         \label{fig:pathV6g_angle}
     \end{subfigure}
     \hfill
        \caption{Test system 1: Bus-voltage observability factors along the dominant path of inter-area mode 1: (a) magnitudes, (b) phases.}
        \label{fig:pathV6g}
\vspace{-0.3cm}
\end{figure}

\begin{figure} [htb!]
\vspace{-0.3cm}
     \centering
     \begin{subfigure}[b]{0.35\textwidth}
         \centering
         \includegraphics[width=\textwidth,trim={0 0.3cm 0 0.3cm},clip]{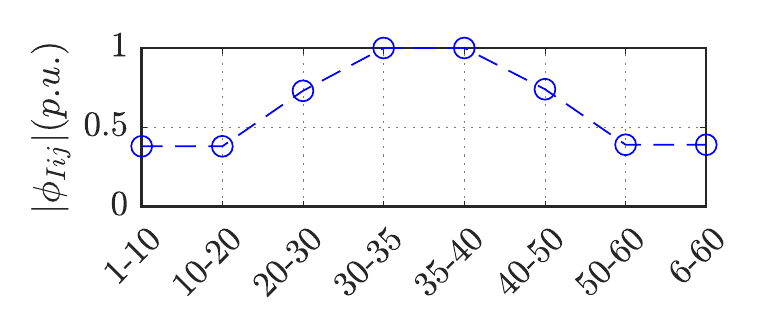}
         \caption{}
         \label{fig:pathI6g_mag}
     \end{subfigure}
     \hfill
     \begin{subfigure}[b]{0.35\textwidth}
         \centering
         \includegraphics[width=\textwidth,trim={0 0.3cm 0 0.2cm},clip]{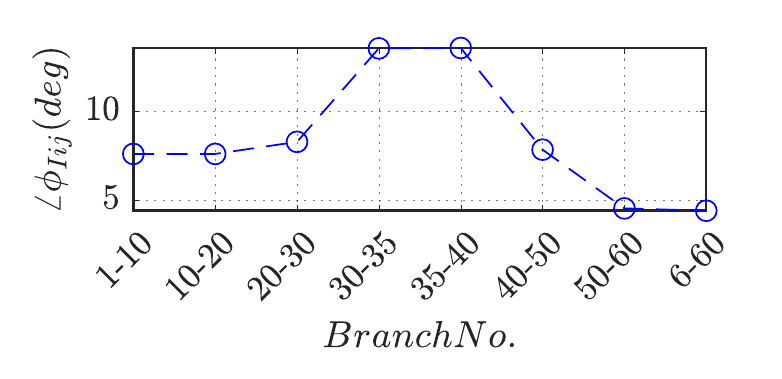}
         \caption{}
         \label{fig:pathI6g_angle}
     \end{subfigure}
     \hfill
        \caption{Test system 1: Branch current observability factors along the dominant path of the inter-area mode 1: (a) magnitudes, (b) phases.}
        \label{fig:pathI6g}
\vspace{-0.3cm}
\end{figure}

Figs. \ref{fig:pathW6g}, \ref{fig:pathV6g} and \ref{fig:pathI6g} show the bus-frequency, bus-voltage and branch-current observability factors along the dominant inter-area oscillation path, respectively. The main characteristics of the dominant path are as follows \cite{chompoobutrgoolIdentificationPowerSystem2013}:

\begin{figure}
\vspace{-0.3cm}
     \centering
     \begin{subfigure}[b]{0.45\textwidth}
         \centering
         \includegraphics[width=\textwidth,trim={0 0.4cm 0 0.25cm},clip]{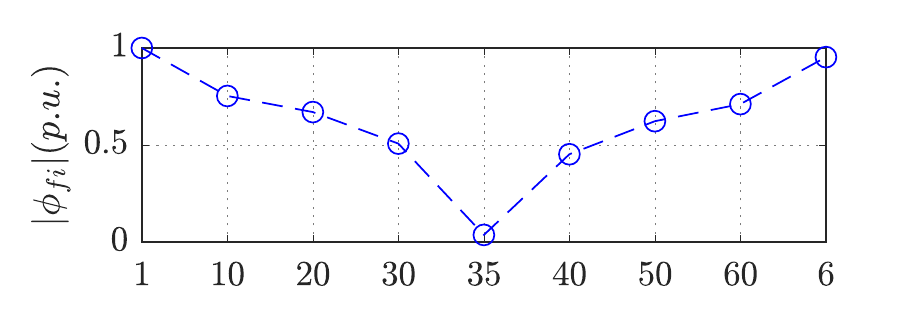}
         \caption{}
         \label{fig:pathW6g_mag}
     \end{subfigure}
     \hfill
     \begin{subfigure}[b]{0.45\textwidth}
         \centering
         \includegraphics[width=\textwidth,trim={0 0.4cm 0 0.1cm},clip]{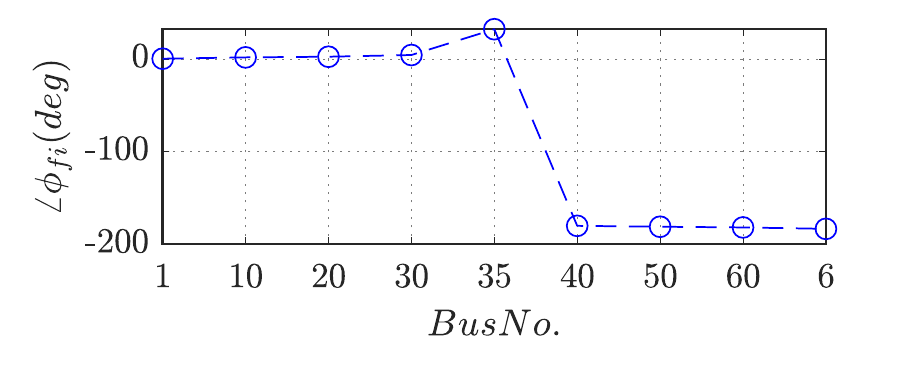}
         \caption{}
         \label{fig:pathW6g_angle}
     \end{subfigure}
     \hfill
        \caption{Test system 1: Bus-frequency observability factors along the dominant path of inter-area mode 1: (a) magnitudes, (b) phases.}
        \label{fig:pathW6g}
\end{figure}

\begin{itemize}
     \item The bus with lowest value of $|\phi_{f_i}|$ (Fig. \ref{fig:pathW6g_mag}) determines the centre of the path, and it is  called the inter-area pivot (or pivot bus, for short) \cite{chompoobutrgoolIdentificationPowerSystem2013} (bus 35, in this example).
     \item The pivot bus divides the path into two groups with opposite phases of $\phi_{f_i}$ (Fig. \ref{fig:pathW6g_angle}). 
     \item The edges of the path have the highest value of $|\phi_{f_i}|$ (Figure \ref{fig:pathW6g_mag}), this confirm that the oscillations are stronger in the edges of the path. In this example, buses 1 and 6 are confirmed as the edges of the path of inter-area mode 1 (see also Fig.~\ref{fig:MS_6g}.)
    \item In test system 1, the highest values of the magnitude of the bus-voltage observability factors, $|\phi_{V_i}|$, correspond to buses close to the centre of the path, while the edges of the path have lower values, as shown in~\cite{chompoobutrgoolIdentificationPowerSystem2013}. Buses close to the centre of the path of this test system also present low values of the phase of bus-voltage observability factors. This observation will deserve further discussion in the next Section.
    \item Branches with higher value of $|\phi_{I_{ij}}|$ (Fig. \ref{fig:pathI6g_mag}) represent those that propagate more the oscillation through the system, and they are often close to the pivot bus. \color{black}
    \item In test system 1, there is no clear pattern in the phases of $\phi_{I_{ij}}$ (Fig. \ref{fig:pathI6g_angle}). They were not analysed in in \cite{chompoobutrgoolIdentificationPowerSystem2013}, probably because these indicators were not so useful to characterise inter-area oscillation paths. 
 \end{itemize}

\FloatBarrier
%\subsection{On the use of bus-frequency or bus-voltage objervability factors to characterise dominant inter-area oscillation paths}
\subsection{Bus-frequency VS. bus-voltage observability factors to characterise dominant inter-area oscillation paths}
\label{sec:path_theory_testsystem2}

Let us consider the system in Fig.~\ref{fig:6g} again but with a load in bus 35 that consumes the power supplied by the generators (test system 2). The data of this new situation are in Section~\ref{sec:Appendix-6g1L} of the Appendix.

%%%%%%%%%%%%%%%%%%%
% DELETED BY AGC
%\begin{figure}[htb!]
%\centering
%\includegraphics[width=0.45\textwidth]{Figures/Section_I/6 generators1L.pdf}
%\caption {Test system 2: Conceptual 6-generator system with one load.}
%\label{fig:6g1L}
%\end{figure}
%%%%%%%%%%%%%%%%%%%%%%%%%%%%%%

\begin{figure}[htb!]
\vspace{-0.3cm}
\centering
\includegraphics[width=0.35\textwidth,trim={0.7cm 0.35cm 0.7cm 0.35cm},clip]{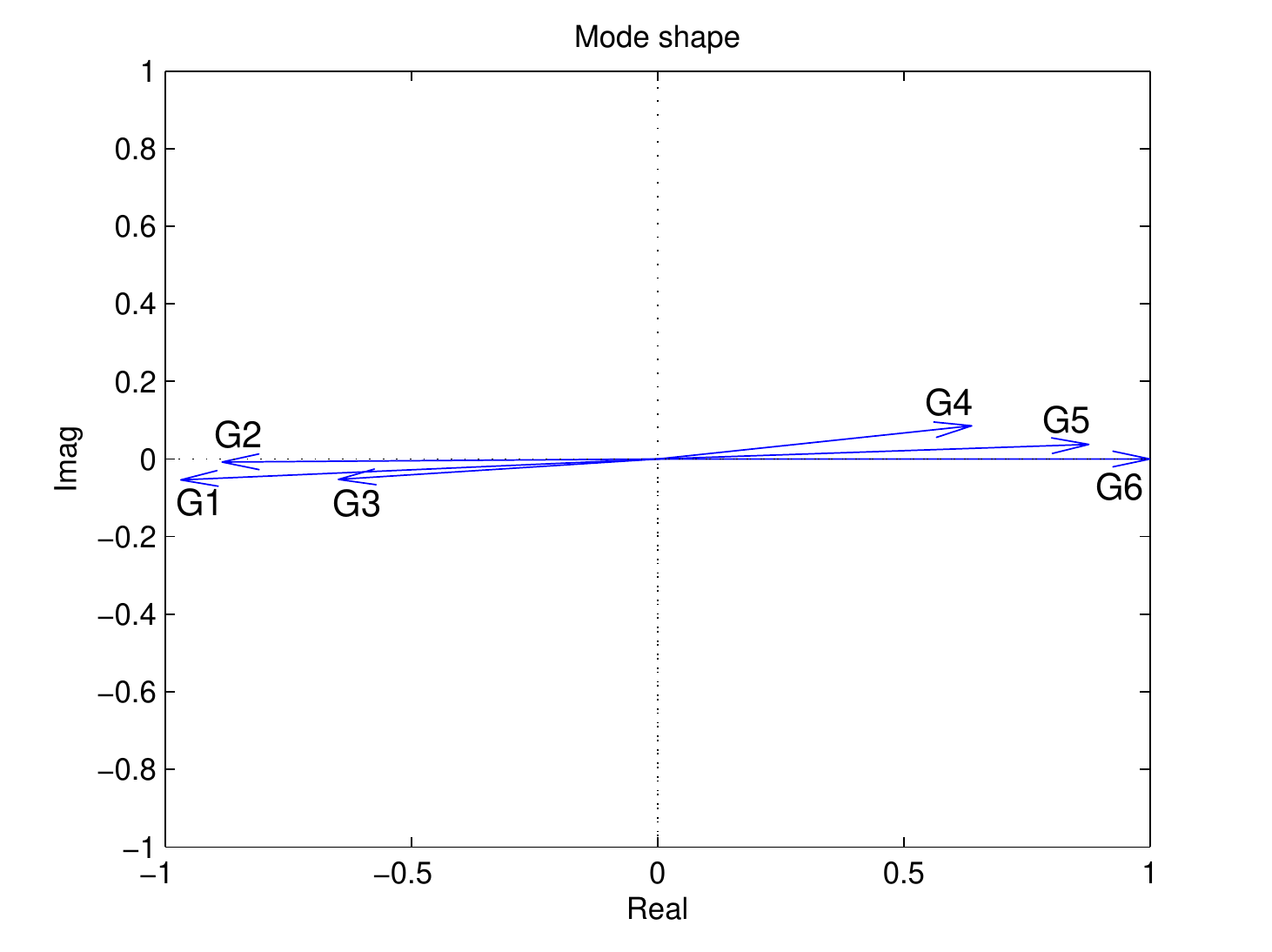}
\caption{Test system 2: Graphical representation of the mode shape of inter-area mode 1.}
\label{fig:MS_6g1L}
\end{figure}

The small-signal analysis of test system 2 shows that this additional load does not strongly impact the main electromechanical characteristics of the system. This system still has a lightly-damped inter-area electromechanical mode (damping coefficient of 11.8 \% and frequency of 0.80 Hz) associated to generators G1, G2 and G3 oscillating against G4, G5 and G6. As for test system 2 (Section~\ref{sec:path_theory}), the largest oscillations are seen in the edges of the two areas (G1 and G6) and the oscillation path includes buses 1, 10, 20, 30, 35, 40, 50, 60 and 6. Mode shapes for test system 2 are plotted in Fig.~\ref{fig:MS_6g1L}.

%%% DELETED BY AGC
%%%%%%%%%%%%%%%%%%%%%%%%%
%\begin{table}[htb!]
%%\resizebox{\columnwidth}{!}
%\centering
%\caption{\label{tab:eigv6g1L} Test system 2: Electromechanical modes.}
%\scalebox{0.65}{
%\begin{tabular}{@{}cccccc@{}}
%\toprule
%No.	&	Real	&	Imag	&	Damp. (\%)	&	Freq. (Hz)	&	Oscillation	\\ \midrule
%1	&	-0.59	&	4.98	&	11.8	&	0.80	&	G1,G2,G3//G4,G5,G6	\\
%2	&	-1.25	&	6.58	&	18.7	&	1.07	&	G2//G1,G3	\\
%3	&	-1.22	&	6.43	&	18.7	&	1.04	&	G3,G6//G1,G2,G4	\\
%4	&	-1.24	&	6.36	&	19.2	&	1.03	&	G4,G3//G1,G2,G6	\\
%5	&	-1.30	&	6.56	&	19.5	&	1.06	&	G5//G4,G6	\\
%	  \bottomrule
%\end{tabular}%
%}
%\end{table}
%%%%%%%%%%%%%%%%%%%%%%%%%%%%

\begin{figure}
\vspace{-0.3cm}
     \centering
     \begin{subfigure}[b]{0.45\textwidth}
         \centering
         \includegraphics[width=\textwidth,trim={0 0.4cm 0 0.3cm},clip]{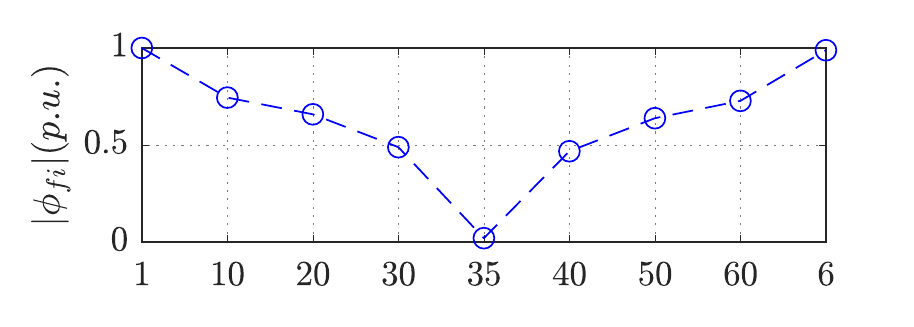}
         \caption{}
         \label{fig:pathW6g1L_mag}
     \end{subfigure}
     \hfill
     \begin{subfigure}[b]{0.45\textwidth}
         \centering
         \includegraphics[width=\textwidth,trim={0 0.4cm 0 0.25cm},clip]{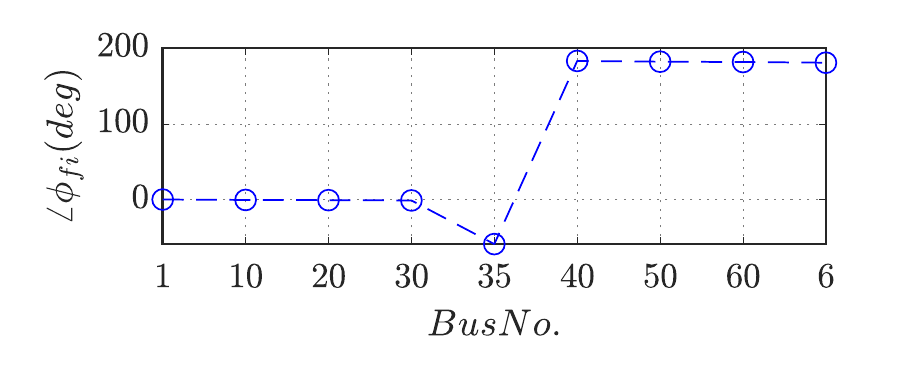}
         \caption{}
         \label{fig:pathW6g1L_angle}
     \end{subfigure}
     \hfill
        \caption{Test system 2: Bus-frequency observability factors along the dominant path of inter-area mode 1: (a) magnitudes, (b) phases.}
        \label{fig:pathW6g1L}
\end{figure}

\begin{figure} [htb!]
\vspace{-0.3cm}
     \centering
     \begin{subfigure}[b]{0.45\textwidth}
         \centering
         \includegraphics[width=\textwidth,trim={0 0.4cm 0 0.3cm},clip]{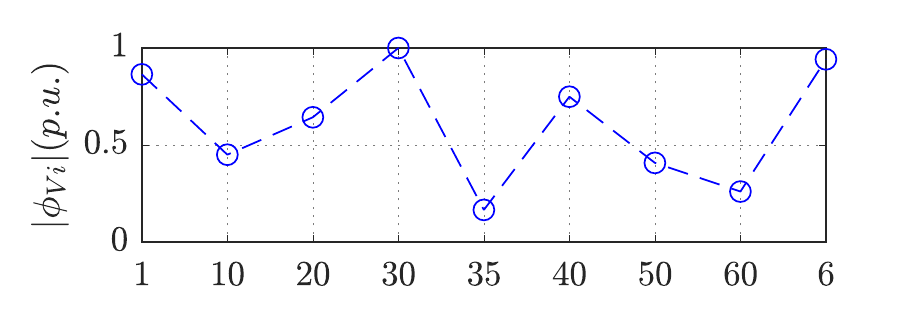}
         \caption{}
         \label{fig:pathV6g1L_mag}
     \end{subfigure}
     \hfill
     \begin{subfigure}[b]{0.45\textwidth}
         \centering
         \includegraphics[width=\textwidth,trim={0 0.4cm 0 0.3cm},clip]{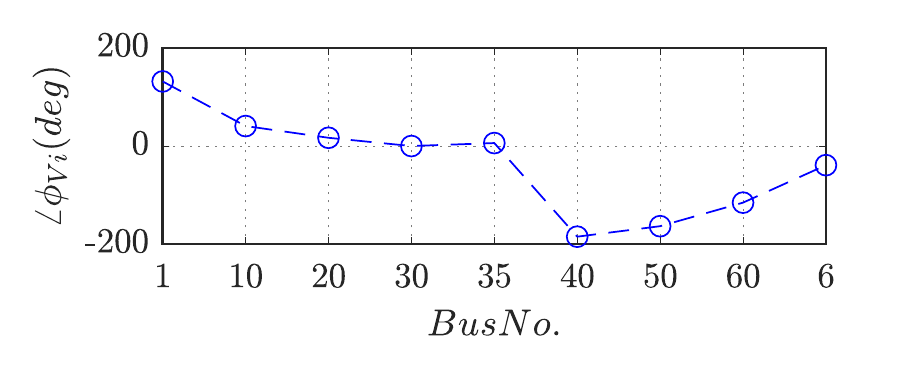}
         \caption{}
         \label{fig:pathV6g1L_angle}
     \end{subfigure}
     \hfill
        \caption{Test system 2: Bus-voltage observability factors along the dominant path of inter-area mode 1: (a) magnitudes, (b) phases.}
        \label{fig:pathV6g1L}
\vspace{-0.3cm}
\end{figure}

\begin{figure} [htb!]
\vspace{-0.1cm}
     \centering
     \begin{subfigure}[b]{0.40\textwidth}
         \centering
         \includegraphics[width=\textwidth,trim={0 0.3cm 0 0.3cm},clip]{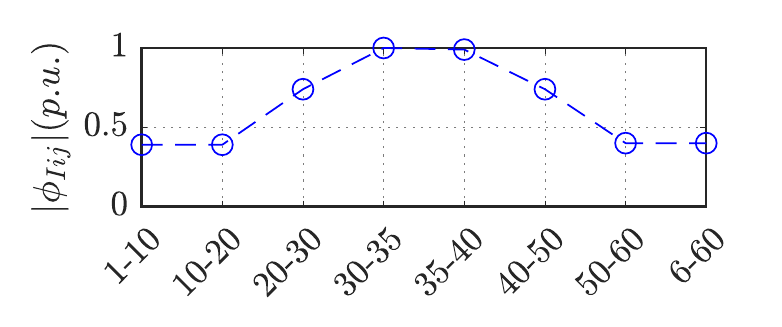}
         \caption{}
         \label{fig:pathI6g1L_mag}
     \end{subfigure}
     \hfill
     \begin{subfigure}[b]{0.40\textwidth}
         \centering
         \includegraphics[width=\textwidth,trim={0 0.3cm 0 0.15cm},clip]{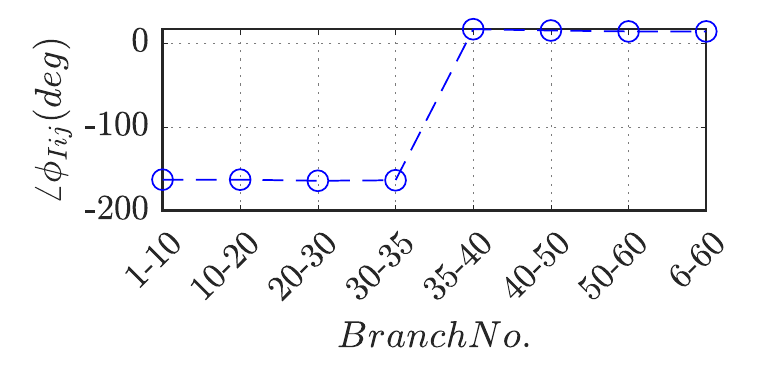}
         \caption{}
         \label{fig:pathI6g1L_angle}
     \end{subfigure}
     \hfill
        \caption{Test system 2: Branch current observability factors along the dominant path of the inter-area mode 1: (a) magnitudes, (b) phases.}
        \label{fig:pathI6g1L}
\vspace{-0.3cm}
\end{figure}

Figs. \ref{fig:pathW6g1L}, \ref{fig:pathV6g1L} and \ref{fig:pathI6g1L} show the bus-frequency, bus-voltage and branch-current observability factors along the dominant inter-area oscillation path in test system 2, respectively.

Results show that:
\begin{itemize}
    \item The conclusions related to the bus-frequency observability factors, $\phi_{f_i}$, are the same as the ones obtained for test system 1. This confirms that bus-frequency observability factors are a robust tool to characterise dominant inter-area oscillation paths and they will be used in this paper.
    \item The conclusions related to the bus-voltage observability factors, $\phi_{V_i}$, are different from those obtained for test system 1. Results depend strongly on the topology and power flows of the system as mentioned without further discussion in~\cite{roucoDampingElectromechanicalOscillations1996}. 
    Hence, bus-voltage observability factors are not considered a robust tool to characterise dominant inter-area oscillation paths.     
    \item The conclusions related to the magnitude of the branch-current observability factors, $|\phi_{I_{ij}}|$, are the same as the ones obtained for test system 1 confirming that bus-frequency observability factors are a robust tool to characterise dominant inter-area oscillation paths and they will be used in this paper.
    \item Finally, the conclusions related to the phase of the branch-current observability factors, $\angle \phi_{I_{ij}}$, are different from the ones obtained for test system 1: the presence of the load affects the pattern of the phases. In fact, they would depend on the direction of the current flows. Hence, the angles of branch-current observability factors are not considered a robust tool to characterise dominant inter-area oscillation paths.
\end{itemize}

\FloatBarrier
\subsection{DC-segmentation of the 6-generator system} \label{sec:DC-s_6g}
\begin{figure}[htb!]
\vspace{-0.3cm}
\centering
\includegraphics[width=0.45\textwidth]{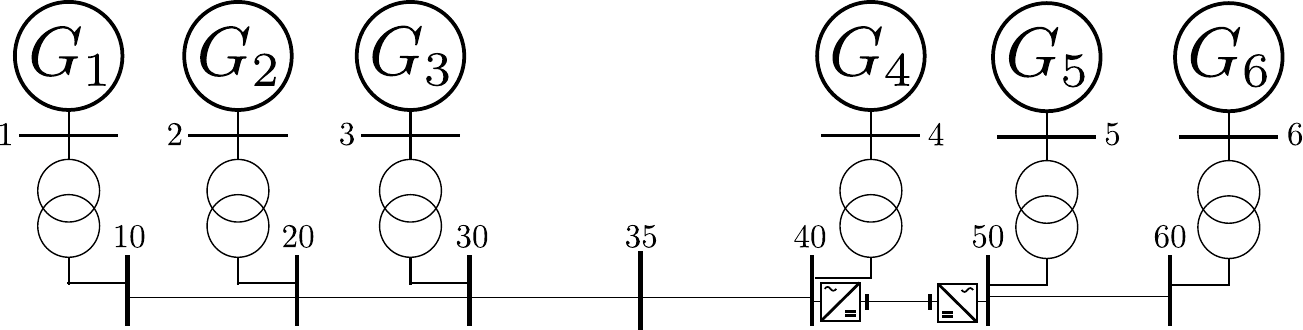}
\caption {Test system 1-DC 2:  6-generator system with DC segmentation 2.}
\label{fig:6gDCs2}
\end{figure}

Once the dominant inter-area oscillation path has been determined, the place where to break it with a DC segment (with VSC-HVDC technology), in order to suppress the critical inter-area mode will be investigated. Test system 1 of Fig. \ref{fig:6g} will be considered and two DC segmentation alternatives will be compared:
\begin{itemize}
    \item DC-segmentation 1: Line 35-40 is replaced by a VSC-HVDC link (Fig. \ref{fig:6gDCs1}) with parameters as in Section~\ref{sec:Appendix-6gDCs} of the Appendix.
    \item DC-segmentation 2: Line 40-50 is replaced by a VSC-HVDC link (Fig. \ref{fig:6gDCs2}) with the same parameters as in the previous case.
\end{itemize}

\begin{figure}[htb!]
\vspace{-0.3cm}
\centering
\includegraphics[width=0.45\textwidth]{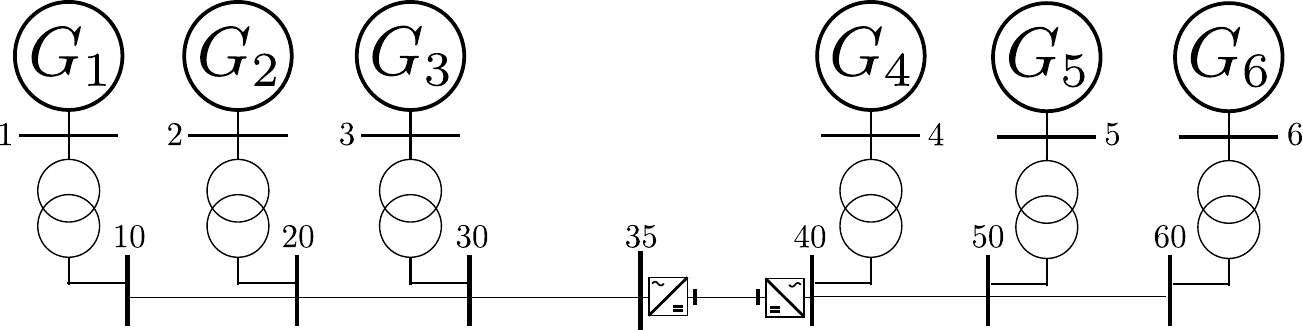}
\caption {Test system 1-DC 1: 6-generator system with DC segmentation 1.}
\label{fig:6gDCs1}
\end{figure}

A small-signal analysis of the system of Fig.~\ref{tab:eigv6gDCs1} gives the electromechanical modes in Table \ref{tab:eigv6gDCs1}(a).
 %\cite{renedoModellingVSCHVDCMultiterminal2019}. 
Results show that the critical inter-area mode has disappeared (compare with Table \ref{tab:eigv6g}) while the remaining four electromechanical modes have similar damping ratios and frequencies to the AC case (Table \ref{tab:eigv6g}). Hence, DC segmentation case 1 improves inter-area oscillation damping, significantly. 

Table \ref{tab:eigv6gDCs1}(b) shows the electromechanical modes DC segmentation case 2. This time, a lightly-damped inter-area mode 1 (G1, G2 and G3 oscillating against G4) is present and, although its damping has improved with respect to most problematic one in  Table~\ref{tab:eigv6g}, it is still worrying. The remaining three local modes have similar damping ratios and frequencies to those in the AC case (Table \ref{tab:eigv6g}). This time the improvement with DC segmentation is not as noticeable as before due to the fact that the DC link has been moved away from the pivot bus of the dominant inter-area oscillation path.

\begin{table}[htb!]
%\resizebox{\columnwidth}{!}
\centering
\caption{\label{tab:eigv6gDCs1} Electromechanical modes of the 6-generator system with DC segmentation}
\scalebox{0.75}{
\begin{tabular}{@{}cccccc@{}}
\toprule
\multicolumn{6}{c}{(a) DC-segmented at line 35-40} \\\midrule
Mode & Real    & Imag   & Damp (\%) & Freq (Hz) & Oscillation \\ \midrule
1	& -1.17	& 6.32	& 18.3	& 1.02	& G4//G6        \\
2	& -1.21	& 6.43	& 18.5	& 1.04	& G3//G1        \\
3	& -1.25	& 6.60	& 18.6	& 1.07	& G2//G1,G3        \\
4	& -1.27	& 6.51	& 19.2	& 1.06	& G5//G4,G6        \\\bottomrule
%\end{tabular}%
%}
%\end{table}
%\begin{table}[htb!]
%\resizebox{\columnwidth}{!}
%\centering
%\caption{\label{tab:eigv6gDCs2} \deleted[id=agc]{Test system 1-DC 2:} Electromechanical modes of the 6 generators system DC-segmented at line 40-50.}
%\scalebox{0.75}{
%\begin{tabular}{@{}cccccc@{}}
%\toprule
\multicolumn{6}{c}{(b) DC-segmented at line 40-50} \\\midrule
Mode & Real    & Imag   & Damp (\%) & Freq (Hz) & Oscillation \\ \midrule
1	& -0.46	& 6.02	& 7.7	& 0.96	& G4//G1,G2,G3   \\		
2	& -1.20	& 6.47	& 18.3	& 1.05	& G3//G1,G4   \\
3	& -1.24	& 6.58	& 18.5	& 1.07	& G1//G2   \\
4	& -1.25	& 6.56	& 18.7	& 1.06	& G5//G6   \\
\bottomrule
\end{tabular}%
}
\end{table}

\section{Proposed algorithm for DC segmentation}\label{sec:DC_segmentation_algorithm}
A meshed AC power system with a critical inter-area mode (with low damping ratio) is considered. The objective of the proposed algorithm is to systematically find a DC segmentation configuration in order to suppress a critical inter-area mode.

A linearised small-signal model of the power system is used to identify the target inter-area mode ($\lambda_{k_{crit}}$) (i.e. the critical inter-area mode, with lowest damping ratio). The proposed algorithm for DC segmentation consists of four steps: %(Fig. \ref{fig:methodo})
\begin{quotation}
\vspace{-0.5cm}
\noindent \begin{itemize}
    \item {\bf Step 1}: Identification of the edges of inter-area oscillation path.
    \item {\bf Step 2}: Identification of the dominant inter-area oscillation path between the two edges.
    \item {\bf Step 3}: DC segmentation: Selection the of AC branch where the inter-area oscillation path will be broken.  
    \item {\bf Step 4}: Check if the present configuration divides the two path edges into two asynchronous AC areas. 
    
    (*) If this occurs, the algorithm stops. Otherwise, return to step 2. 
   \item {\bf Finally}: Once the algorithm stops, each AC line selected for DC segmentation will be replaced by a VSC-HVDC link.
 \end{itemize}
\end{quotation}

% In step 4, we cannot speak about DC-segmentation configuration, since the system will be DC segmented only once there won't be any link between E1 and E2. 
% old version of step 4: Step 4: Check if the DC-segmentation configuration divides the two path edges into two asynchronous AC areas. If this occurs, the algorithm stops. Otherwise, return to step 2. \\

The algorithm will use the information provided by the modal analysis of the linearised model of the power system which has to be executed only once at the beginning and it is not repeated at every iteration of the loop. Hence, the computational burden of the algorithm is low. The algorithm uses the following information:
\begin{itemize}
    \item Mode shapes of the target inter-area mode participating in the speeds of the generators ($v_{i,k_{crit}}$).
    \item Frequency observability factors of the target inter-area mode of all buses ($\phi_{f_{i,k_{crit}}}$).
    \item Magnitude ($|\phi_{I_{ij},k_{crit}}|$) of the current observability factors of the target inter-area mode of all branches ($\phi_{I_{ij},k_{crit}}$).
\end{itemize}
For the sake of clarity, the sub-index of the target inter-area mode will be removed from the indicators above in the future (i.e., $\phi_{f_{i}}$ will be used instead of $\phi_{f_{i,k_{crit}}}$).
\subsection{Step 1: Identification of the edges of inter-area oscillation path} \label{sec:Step1}
The bus of the first edge of the path will be called $E1$ and it contains a synchronous generator ($GE1$). The bus of the other edge will be called $E2$ and it contains another synchronous generator ($GE2$). 
The edges of the path are obtained as follows:

 \begin{itemize}
    \item The first edge of the inter-area oscillation path (bus $E1$) will be determined by the generator $GE1$ with the largest magnitude of the inter-area mode shape: $|v_{GE1}|=\max_i|v_{i}|$. 
    \item The second edge of the inter-area oscillation path (bus $E2$) will be determined by the generator $GE2$ with the largest magnitude of the inter-area mode shape among the generators which are oscillating with a phase greater than $90^\circ$ with respect to generator $GE1$. In other words, generator $GE2$ is the one that satisfies $|v_{GE2}|=\max_i|v_{i}|$, among those that satisfy $|\angle v_{GE1}- \angle v_{i}|>90^\circ$. 
    \item All generators $i$ with $|v_{i}|>0.1$ pu and $|\angle v_{GE1}-\angle v_{i}|<45^\circ$ will belong to the coherent group of generators associated with edge $E1$.
    \item All generators $i$ with $|v_{i}|>0.1$ pu and $|\angle v_{GE2}- \angle v_{i}|<45^\circ$ will belong to the coherent group of generators associated with edge $E2$.
 \end{itemize}
 
Notice that there could be inter-area modes with more than two coherent groups of generators. This algorithm will focus only on the first two groups.

\subsection{Step 2: Identification of the dominant inter-area oscillation path between the two edges} \label{sec:Step2bis}
The algorithm presented here revisits the idea presented in~\cite{chompoobutrgoolIdentificationPowerSystem2013} but includes some additions needed to make the process completely automatic. The main additions are highlighted in Section~\ref{sec:Appendix-inter_area_oscillation_path_comparison} of the Appendix. 

The target inter-area oscillation path will go from $E1$ to $E2$. In order to tackle meshed systems, where more than one propagation path is possible for the target inter-area mode, let us add the following definitions:
\begin{itemize}
    \item $A_{path,ip}$ is the set of branches of the system ($L_{ij}$) that belong to path-$ip$.
    \item $A_{path,bus,ip}$ is the set of buses of the system ($i$) that belong to path-$ip$.
    \item $A_{EX,ip}$ is the set of branches of the system ($L_{ij}$) that are excluded to be selected as part of path-$ip$ during the process. These branches could be branches that already belong to the path, or branches that belong to an unfeasible radial path that has already been broken to a previous DC segment, as will be explained later.
    %\item $A_{PATHS}$ is defined as the set of branches of the system ($L_{ij}$) that belong to any of the inter-area oscillations paths: $A_{PATHs}=A_{path,1} \cup A_{path,2} \cup \ldots$
    \item $A_{L,i}$ is the set of branches that are connected to a certain bus $i$.  
    \item $A_{DC,segs}$ is the set of branches of the system ($L_{ij}$) that have been selected for DC segmentation during the the execution of the algorithm.
\end{itemize}
Sets $A_{path,ip}$, $A_{path,bus,ip}$ and $A_{EX,ip}$ are empty at the beginning of Step 2. Set $A_{DC,segs}$ is empty when the algorithm starts. Notice that all inter-area oscillation paths will start in edge $E1$ and will end in edge $E2$. 

Recalling the definition of ``pivot bus'' ($PB$) in Section~\ref{sec:path_theory} ($PB$ is the bus of the path with the minimum value of the magnitude of the bus-frequency observability factor: $|\phi_{f_{PB}}| = \min_{i \in A_{path,bus,jp}}|\phi_{f_{i}}|$), the dominant inter-area oscillation path-$ip$ can be split in two parts:

\begin{itemize}
    \item Descending sub-path: From $E1$ to $PB$.
    \item Ascending sub-path: From $PB$ to $E2$.
\end{itemize}

The descending sub-path will tear along the branches with the highest current observability factor linking buses with a decreasing frequency observability factor. The ascending sub-path will tear along the branches with the highest current observability factor linking buses with increasing frequency observability factor.
The end of the descending sub-path will be established by detecting the PB which splits the path in two parts with opposite phase in the frequency observability factors (Section~\ref{sec:path_theory}).

When moving along the descending sub-path, if the bus current observability factor increases, it means that:
\begin{itemize}
    \item either that the ascent has started,i.e., the first bus of the ascent ($A$, for short) has been reached;
    \item or the line selected is going "backwards",i.e., in the direction of E1. In this case, this line must be suppressed from the sub-path.
\end{itemize}

If the ascent has been reached, the pivot bus has been passed. Thus, bus $A$ is the first bus of the path that respects:
\begin{equation}\label{eq:DC_segmentation_path_AB}
|\angle \phi_{f_{A}}- \angle \phi_{f_{E1}}|>90^\circ
\end{equation}

Although bus $A$ is part of the ascending sub-path, it will actually be identified during the descending sub-path.

Before starting the process, include the AC branches that have been selected to be replaced by DC segments ($L_{ij} \in A_{DC,segs}$) into the set of the excluded branches of path-$ip$ ($A_{EX,ip}$).

\subsubsection{Descending sub-path} \label{sec:Step2_descent}
The descending sub-path is determined as follows:
%\begin{quotation}
\noindent \begin{algorithmic}
%\PROCEDURE
\STATE {\bf Step a}: It starts at the first edge of the inter-area oscillation path (bus $i=E1$).
\STATE {\bf Step b}: Feasibility check: Consider all branches $L_{ij}$ connected to bus $i$: $L_{ij} \in A_{L,i}$. Check if those branches have not been excluded to be selected as the next line of the path ($L_{ij} \in A_{EX,ip}$). 
\bindent
\IF{there are no feasible candidates ($A_{L,i} \cap A_{EX,ip}^C= \emptyset$),}
\STATE Exclude the last branch of the path, go back to the previous bus of the path (i.e., include $L_{i-1,i}$ set $A_{EX,ip}$ and put $i=i-1$) and repeat {\bf Step b}
\ELSE 
\STATE Continue
\ENDIF
\eindent
\STATE The feasibility check has two purposes: excluding branches that are already part of the path, and excluding branches that belong to a radial part of the system that has already been broken by a previous DC segment. 
\STATE {\bf Step c}: Consider all the feasible branches $L_{ij}$ connected to bus $i$: $L_{ij} \in A_{L,i} \cap A_{EX,ip}^C$ and choose the one with the largest magnitude of the branch-current observability factor $|\phi_{I_{ij}}|$. The next bus selected will be called $jp$ and thus the selected line will be called $L_{i,jp}$.
\bindent
\IF{in $L_{i,jp}$ the bus-frequency observability factor decreases (i.e., $|\phi_{f_{ip}}|<|\phi_{f_{i}}|$),} 
    \STATE  we are still on the descent, branch $L_{i,jp}$ and bus $jp$ can be added to the path and $L_{i,jp}$ is included into $A_{path,ip}$ and bus $jp$ is included into $A_{path,bus,ip}$. At this point, branch $L_{i,jp}$ is part of path-$ip$ and cannot be selected any more. Hence, it is included into the set of excluded lines $A_{EX,ip}$. Finally, put $i=jp$ and return to {\bf Step b}.
    \ELSE 
    \STATE 
    Go to {\bf Step d}
    \ENDIF
\eindent
\STATE {\bf Step d}: Check if bus ip is part of the ascending sub-path. 
\bindent
\IF{If $|\angle \phi_{f_{E1}}- \angle \phi_{f_{jp}}|>90^\circ$} 
\STATE bus $jp$ is the first one of the ascent: $A=jp$. Branch $L_{i,jp}$ and bus $jp$ can be added to the path. The descending sub-path can be exited and the ascending sub-path started.
\ELSE
\STATE Bus $jp$ correspond to a dead end, exclude branch $L_{i,jp}$ of the path, go back to the previous bus of the path, and repeat {\bf Step b}.
\ENDIF
\eindent
%\ENDPROCEDURE
\end{algorithmic}
%\end{quotation}

\subsubsection{Ascending sub-path} \label{sec:Step2_ascent}
The ascending sub-path is determined as follows:
\begin{algorithmic}
\STATE {\bf Step a}: It starts at the bus $A$ identified during the descending sub-path.
\STATE {\bf Step b}: Feasibility check:
\bindent
\IF{there are feasible branches $L_{ij}$ connected to bus $i$} 
\STATE Continue
\ELSE 
\STATE (i.e. $A_{L,i} \cap A_{EX,ip}^C= \emptyset$), include branch $L_{i-1,i}$ into set $A_{EX,ip}$, put $i=i-1$ and repeat {\bf Step b.}
\ENDIF
\eindent
\STATE {\bf Step c}: Consider all the feasible branches $L_{ij}$ connected to bus $i$ (i.e., $L_{ij} \in A_{L,i} \cap A_{EX,ip}^C$).  From those branches, choose the one with largest magnitude of the branch-current observability factor $|\phi_{I_{ij}}|$. The next bus selected will be called $jp$.
\bindent
\IF{if this branch the magnitude of the bus-frequency observability factor increases (i.e., $|\phi_{f_{ip}}|>|\phi_{f_{i}}|$),}
\STATE bus $jp$ will  the next bus of the path. Branch $L_{i,jp}$ is included into $A_{path,ip}$ and $A_{EX,ip}$, and $jp$ is included into $A_{path,bus,ip}$. 
\ELSE
\STATE $L_{i,jp}$ correspond to a dead end, exclude it from the path, go back to the previous bus of the path, and repeat {\bf Step b}. 
\ENDIF
\eindent
\STATE {\bf Step d}: Check if the second edge of the path has been reached 
\bindent
\IF{$jp=E2$,} 
\STATE stop.
\ELSE 
\STATE put $i=jp$ and return to {\bf Step b}.
\ENDIF
\eindent
\end{algorithmic}

\subsection{Step 3:  Selection the AC branch where the inter-area oscillation path will be broken} \label{sec:Step3}
Once the inter-area oscillation path has been identified, the next step is to select the AC line which will be replaced by a VSC-HVDC link at the end of the algorithm. As discussed in Section \ref{sec:DC-s_6g}, DC segmentation is more effective when placed close to the centre of the inter-area oscillation path. Therefore, the two lines of the path connected to the $PB$ will be the candidates for segmentation. However, the line with the highest value of the magnitude of the branch-current observability factor is selected, because this line will be the one with the most intense inter-area oscillation. The selected AC line, is then included into set $A_{DC,segs}$.

\subsection{Step 4: Is the system DC-segmented?} 
Each iteration of the algorithm (Steps 2 and 3) determines a path of the target inter-area oscillation (path $ip$) and the AC branch to be replaced by a DC segment. This does not guarantee dividing the system into two asynchronous areas, as the DC segments could contain parallel AC paths. Step 4 scans the grid looking for a continuous connection between E1 and E2 through an AC path. If an AC connection is found, the algorithm goes back to Step 2 and repeats the process until no such connection is found and the algorithm can be terminated.

Once the algorithm has stopped, each AC line selected for DC segmentation ($L_{ij} \in A_{DC,segs}$) is replaced with a VSC-HVDC link with the same nominal apparent power.

\subsection{Illustrative example} \label{sec:application_6g}
The proposed algorithm is now illustrated in the 6-generator system of Fig. \ref{fig:6g} (test system 1). Although it is a simple power system with radial configuration, it is useful to help to understand the proposed algorithm. The algorithm has been implemented in Matlab and linked with a tool box for small-sinal stability analysis (SSST) \cite{roucoSmallSignalStability2002}.

The results of the algorithm when applied to test system~1 (Fig. \ref{fig:6g}) can be summarized as follows:
\begin{enumerate}
    \item {\bf Step 1}: The first edge of the inter-area oscillation path is bus 1 (E1=1). The coherent generators of this group are generators G1, G2 and G3. The second edge of the inter-area oscillation path is bus 6 (E2=6). The coherent generators of this group are generators G4, G5 and G6.
    \item {\bf Step 2}:
    \begin{itemize}
        \item The descending sub-path is 1-10-20-30-35. \\
        Bus 40 has been identified as the first bus of the ascending sub-path ($A=40$). \\
        Bus 35 has been identified as the pivot bus of the path ($PB=35$).
        \item The ascending sub-path is 40-50-60-6.
    \end{itemize}
    Hence, the propagation path is 1-10-20-30-35-40-50-60-6.
    \item {\bf Step 3}: AC line 35-40 is selected for DC segmentation.
    \item {\bf Step 4}: The system has been divided into two asynchronous areas and stops. Line 35-40 is replaced by a VSC-HVDC link with the same nominal apparent power as the AC line replaced. 
\end{enumerate}

Notice that the DC-segmentation obtained here is the one of Fig.~\ref{fig:6gDCs1} which succeeded in suppressing the critical inter-area mode, as shown in Section~\ref{sec:DC-s_6g}. 

\FloatBarrier

\section{Case study and results}\label{sec:results}
\begin{figure*}[!htb]
\vspace{-0.3cm}
\centering
\includegraphics[width=0.9\textwidth]{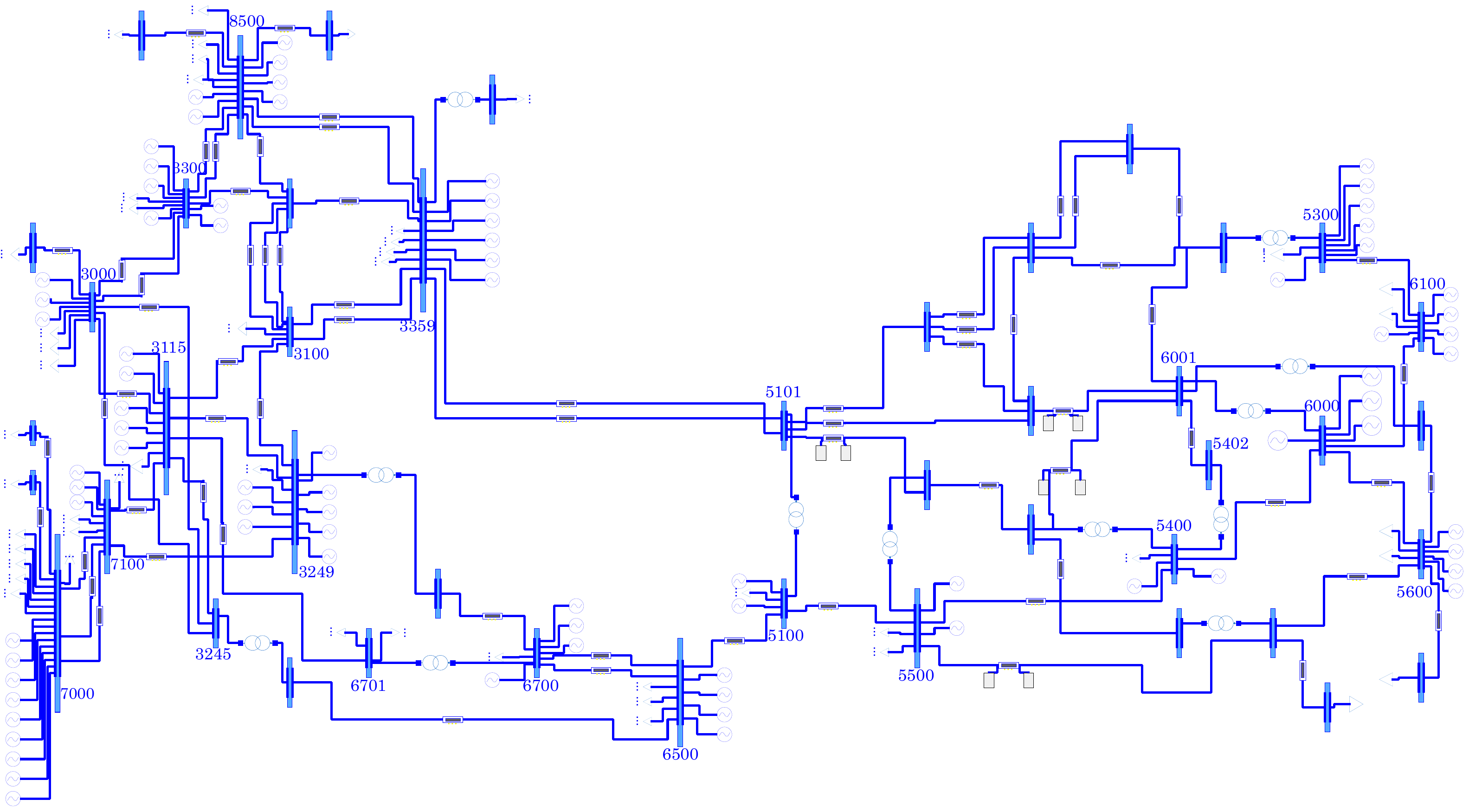}
\caption {N44 test system under Dymola.}
\label{fig:N44system}
\end{figure*}

The proposed algorithm has also been applied to the Nordic 44 test system (Fig.~\ref{fig:N44system}) which is a representation of the interconnected grids of Norway, Sweden and Finland. It is based on previous models developed at The Norwegian University of Science and Technology (NTNU) \cite{jakobsenNordic44Test2018}. It has been initially implemented within iTesla project as an application example of the OpenIPSL library, implemented in the Modelica language \cite{vanfrettiITeslaPowerSystems2016, baudetteOpenIPSLOpenInstancePower2018}. The version used in this paper is the one updated by the ALSETlab team. In this paper, the simulations of the Nordi 44 system are carried out using the Dymola environment. OpenIPSL can be used for non-linear electromechanical-type simulation, but also for small-signal stability analysis, by using a numerical linearisation of the system. The information about the scenario considered is provided in Section~\ref{sec:Appendix-N44_AC_base_case} of the Appendix .

\begin{figure*}[!htb]
\vspace{-0.3cm}
\centering
\includegraphics[width=0.9\textwidth]{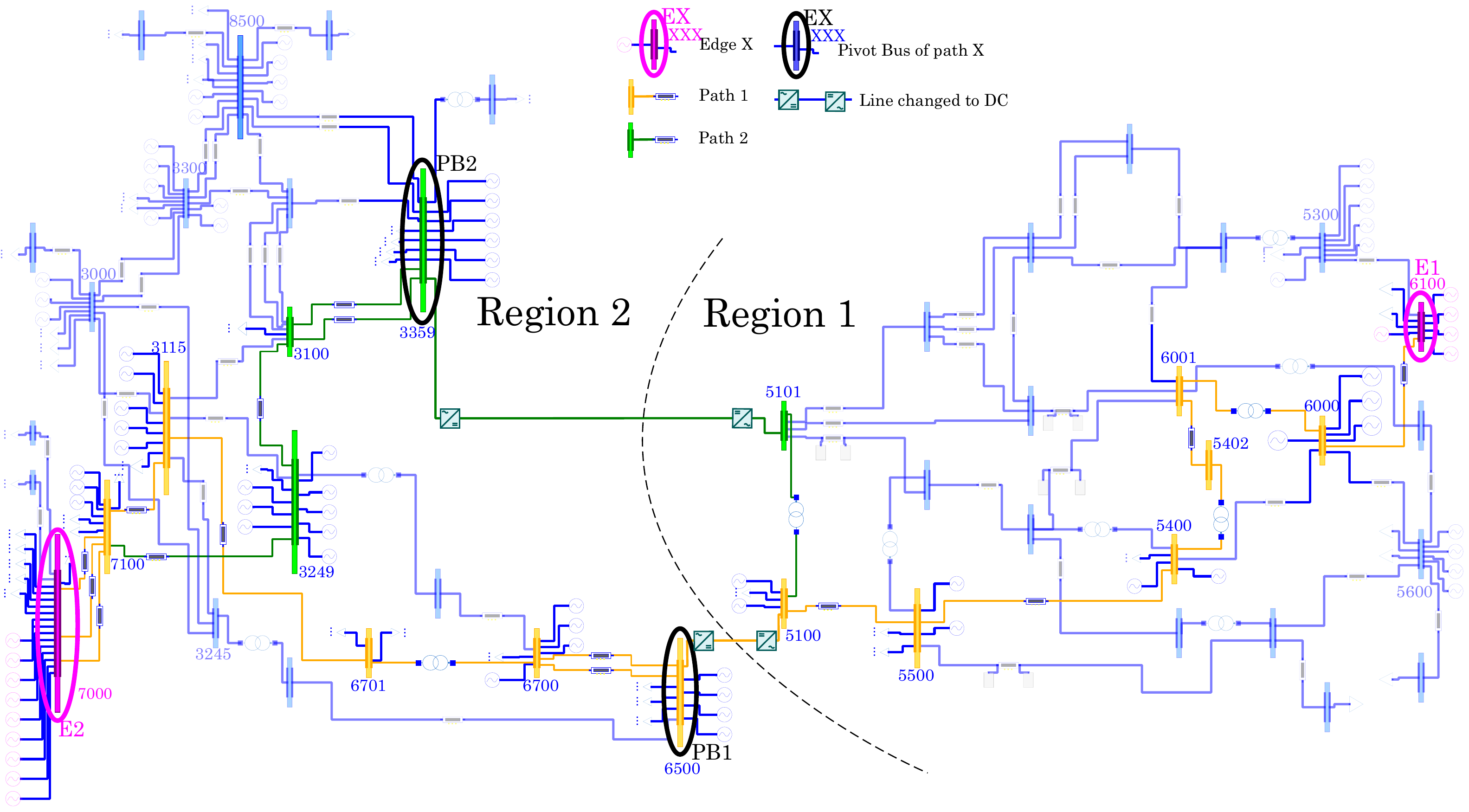}
\caption {DC-segmented N44 test system under Dymola.}
\label{fig:N44systemDCs}
\vspace{-0.3cm}
\end{figure*}

\subsection{Application of the proposed algorithm}
The result of the algorithm applied to the Nordic 44 system is depicted in Fig. \ref{fig:N44systemDCs}. All buses and lines of interest for the discussion that follows are presented in colours.

Table \ref{tab:eigvN44_v1} shows the poorly damped electromechanical modes of Nordic 44 test system, (damping ration under 20\%). They have been obtained with OpenIPSL. Mode 1 (damping of 1.85\% and frequency of 0.39 Hz) is selected as the target inter-area mode for the DC segmentation algorithm because it has the lowest damping ratio. In inter-area mode 1, generators of the South of Norway (named region 1 in Fig.\ref{fig:N44systemDCs}) are oscillating against most of the remaining generators of the system.

\begin{table}[htb!]
%\resizebox{\columnwidth}{!}
\centering
\caption{\label{tab:eigvN44_v1} Electromechanical modes of Nordic 44 system with low damping ratio.}
\scalebox{0.80}{
\begin{tabular}{@{}ccc}
\toprule
    &   \multicolumn{2}{c}{AC base case}  \\
N0. &   Damp (\%) & Freq (Hz) \\ \midrule
1	&	1.85	&	0.39		\\
2	&	5.45	&	0.83		\\
3	&	12.22	&	0.54		\\
4	&	12.11	&	0.75		\\
5	&	11.68	&	0.88		\\
6	&	13.12	&	0.98		\\
7	&	12.12	&	1.07		\\
%8	&	-	&	-	&	12.06	&	1.12	\\
8	&	13.57	&	1.23		\\
9	&	15.69	&	1.10		\\
10	&	15.17	&	1.88		\\
11	&	16.69	&	1.77		\\
\bottomrule
\end{tabular}%
}
\end{table}

The proposed algorithm is implemented in Matlab+SSST. This tool is used, because the information needed for the implementation of the algorithm (e.g., mode shapes and observability factors) is not provided by the linearised model of OpenIPSL.

Since the toolbox SSST does not accept more than one generator connected to a bus, generators on the same bus in the original system have been aggregated into a single unit. Likewise, if a pair of buses were connected by more than one circuit (e.g. buses 7100 and 7000 of Fig. \ref{fig:N44system}), this circuits were aggregated into a single line, before running the algorithm.
%Javier: I ahve a doubt between using the term "Dymola". Question: OpenIPSL uses Modelica language and Dymola environment?
% Should we use "the Dymola model" or the "OpenIPSL" model?

% The details of the aggregations are presented in Appendix \ref{sec:Appendix-N44_aggregation}.

\subsubsection{Step 1: Identification of the path edges}
\begin{figure}[htb!]
\vspace{-0.3cm}
\centering
\includegraphics[width=0.3\textwidth,trim={0.7cm 0.65cm 0.7cm 0.45cm},clip]{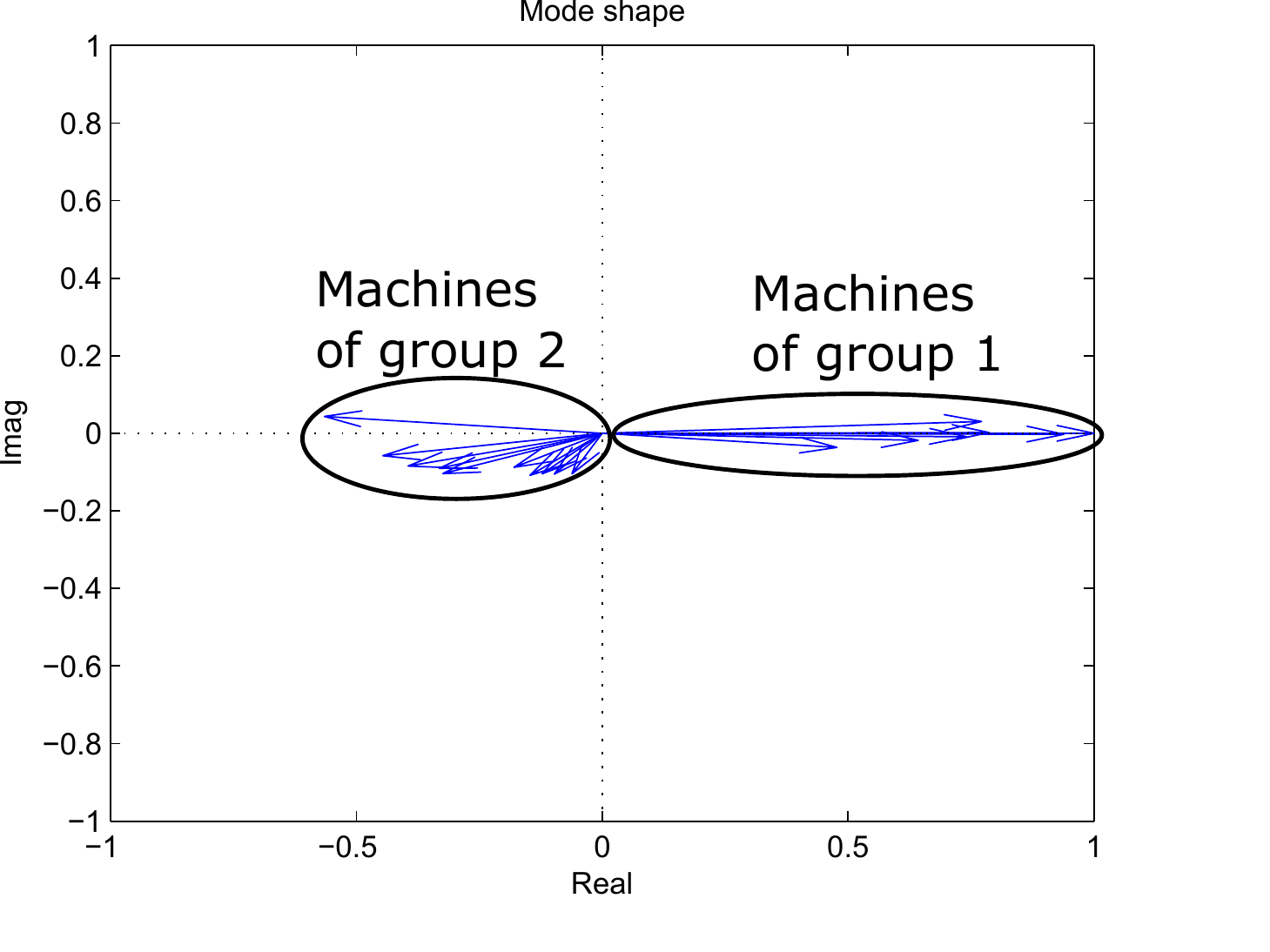}
\caption{Mode shape of the Nordic 44 test system in mode 1.}
\label{fig:MS_N44}
\vspace{-0.3cm}
\end{figure}

Fig.\ref{fig:MS_N44} shows the generator mode shapes of the system.

The first edge of the inter-area-oscillation path is bus 6100 (E1=6100). The coherent generators of this group are 6100, 5300, 5600, 6000, 5400, 5500 and 5100. 
The second edge of the inter-area-oscillation path is bus 7000 (E2=7000). The coherent generators of this group are 7000, 7100, 3249, 3115, 6700, 3000, 3245 and 3300.

In Fig.\ref{fig:N44systemDCs}, generators 3359 and 7000 (the path edges) are coloured in pink, group 1 corresponds to the generators of region 1 while group 2 corresponds to the generators of region 2 (with the exception of generators 6500, 8500 and 3359 that are not in those groups).
  
\subsubsection{Step 2, 3 and 4 first iteration: identification and breaking of the first dominant inter-area path}
{\bf Step 2}: The identified path (coloured in orange in Fig.~\ref{fig:N44systemDCs}) consists of buses 6100 (E1), 6000, 6001, 5402, 5400, 5500, 5100, 6500, 6700, 6701, 3115, 7100, 7000 (E2). 

Bus 6500 was identified as the pivot bus of the path ($PB=35$) and bus 6700 was identified as the first bus of the ascending sub-path ($A=40$).

\begin{figure}
     \centering
     \begin{subfigure}[b]{0.45\textwidth}
         \centering
         \includegraphics[width=\textwidth,trim={0 0.35cm 0 0.35cm},clip]{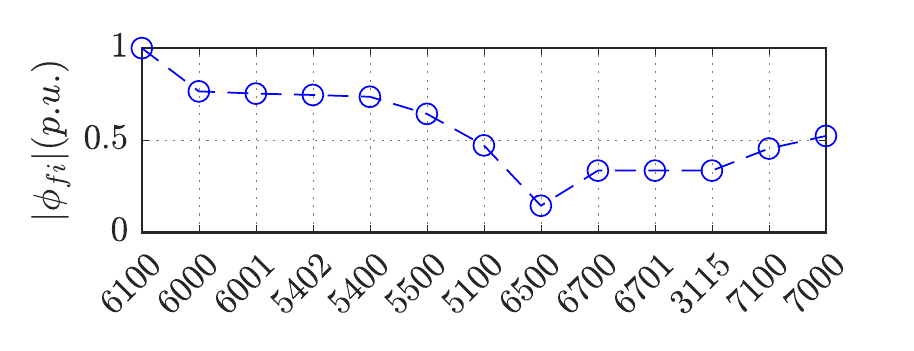}
         \caption{}
         \label{fig:N44path1Sw_mag}
     \end{subfigure}
     \hfill
     \begin{subfigure}[b]{0.45\textwidth}
         \centering
         \includegraphics[width=\textwidth,trim={0 0.35cm 0 0.25cm},clip]{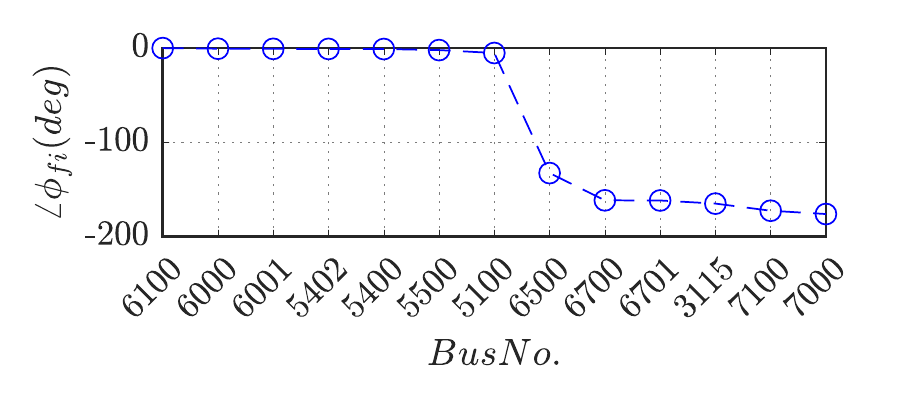}
         \caption{}
         \label{fig:N44path1Sw_angle}
     \end{subfigure}
     \hfill
        \caption{Bus frequency observability factor along the first dominant path (path 2) in the Nordic 44 test system: (a) magnitude and (b) phase.}
        \label{fig:N44path1Sw}
\vspace{-0.3cm}
\end{figure}

\begin{figure}[htb!]
\centering
\includegraphics[width=0.45\textwidth,trim={0 0.4cm 0 0.25cm},clip]{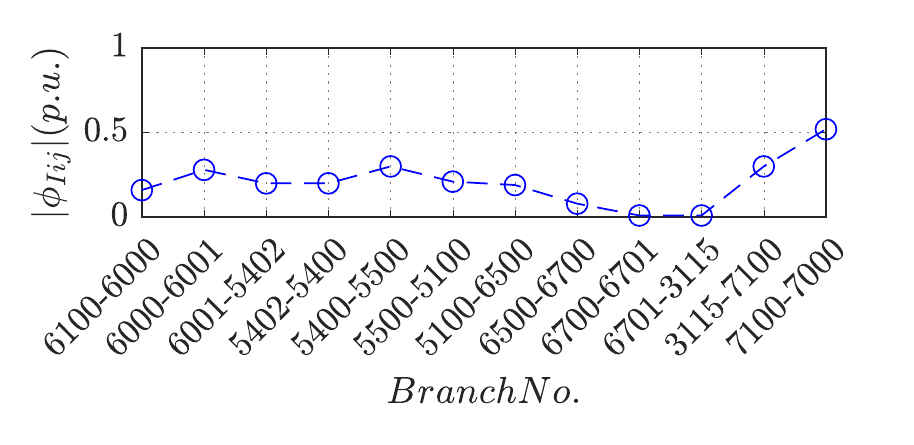}
\caption{Magnitude of branch current observability factor along the first dominant path (path 1) in the Nordic 44 test system.}
\label{fig:N44path1SI}
\end{figure}

Fig. \ref{fig:N44path1Sw} shows the bus frequency observability factors along this path and confirms the observations made in the Section~\ref{sec:path_theory}. Fig. \ref{fig:N44path1SI} shows the branch current observability factors along this path.

{\bf Step 3}: Line 5100-6500 is selected for DC segmentation.

{\bf Step 4}: With the first DC segment, the system is not divided into two asynchronous AC systems. Thus, the algorithm goes back to step 2. 

\subsubsection{Step 2, 3 and 4 second iteration: identification and breaking of the second dominant inter-area path}
{\bf Step 2}:  The identified path comprises buses 6000 (E1), 6001, 5402, 5400, 5500, 5100, 5101, 3359, 3100, 3249, 7100 (E1). Several buses are also in the previous path. In Fig. \ref{fig:N44systemDCs}, the part of path 2 that differs from path 1 iscoloured in green.

Bus 3359 was identified as the pivot bus of the path ($PB=35$), and bus 3100 was identified as the first bus of the ascending sub-path ($A=40$).

\begin{figure}
\vspace{-0.3cm}
     \centering
     \begin{subfigure}[b]{0.45\textwidth}
         \centering
         \includegraphics[width=\textwidth,trim={0 0.35cm 0 0.3cm},clip]{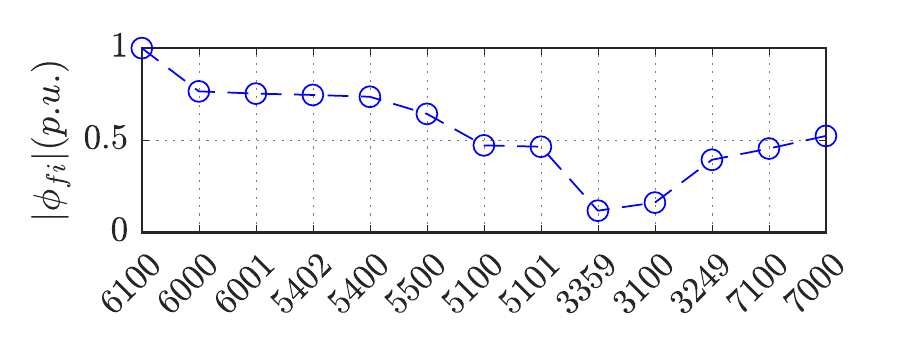}
         \caption{}
         \label{fig:N44path2Sw_mag}
     \end{subfigure}
     \hfill
     \begin{subfigure}[b]{0.45\textwidth}
         \centering
         \includegraphics[width=\textwidth,trim={0 0.35cm 0 0.25cm},clip]{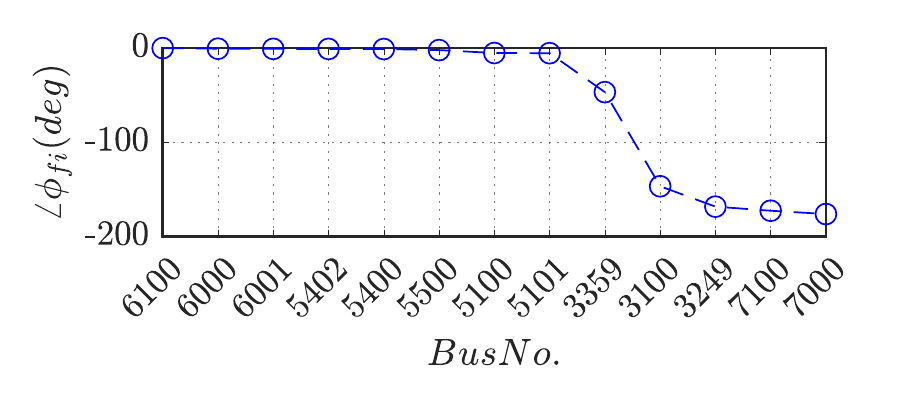}
         \caption{}
         \label{fig:N44path2Sw_angle}
     \end{subfigure}
     \hfill
        \caption{Bus frequency observability factor along the second dominant path (path 2) in the Nordic 44 test system: (a) magnitude and (b) phase.}
        \label{fig:N44path2Sw}
\end{figure}

\begin{figure}[htb!]
\vspace{-0.3cm}
\centering
\includegraphics[width=0.45\textwidth,trim={0 0.4cm 0 0.25cm},clip]{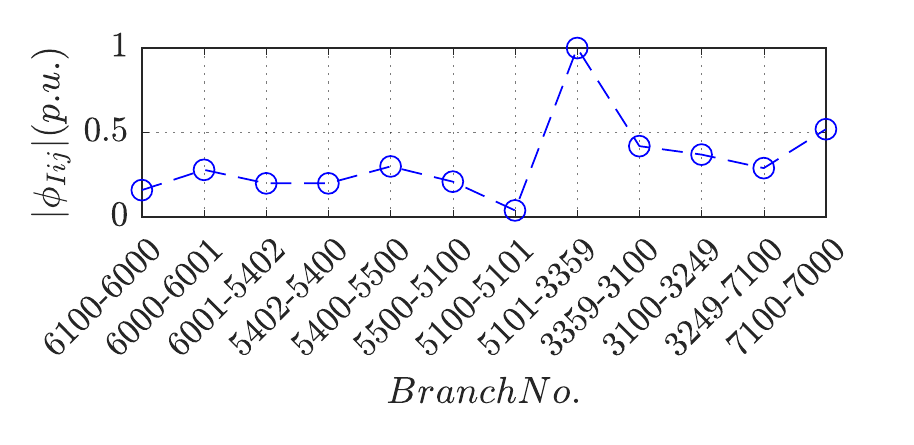}
\caption{Magnitude of branch current observability factor along the second dominant path (path 2) in the Nordic 44 test system.}
\label{fig:N44path2SI}
\vspace{-0.3cm}
\end{figure}

Fig. \ref{fig:N44path2Sw} shows the bus frequency observability factors along this path and confirms the observations made in the Section \ref{sec:path_theory}. Fig. \ref{fig:N44path2SI} shows the branch current observability factors along this path.

{\bf Step 3}: Line 3359-5101 is selected for DC segmentation.

{\bf Step 4}: The system has now been split into two AC areas.

%%% AGC: i HAVE CHANGED THE TITLE A BIT
\subsubsection{Algorithm termination}
Lines 5100-6500 and 3359-5101 were replaced by VSC-HVDCs links of 800 MVA (path 1) and 3500 MVA (path 2), to complete DC-segmentation. 

The original AC system was split in two AC clusters, one with all the generators of group 1 and another one with all the generators of group 2 (regions 1 and 2, respectively, in Fig.~\ref{fig:N44systemDCs}). 

%\FloatBarrier
\subsection{Validation of the algorithm}\label{results_validation}
The effects of the DC-segmentation proposed were investigated in the system in using Dymola. VSC-HVDC links were modelled as proposed in \cite{coleGeneralizedDynamicVSC2010a} and models were implemented in Modelica language to be tested in combination with the OpenIPSL library. Two cases were compared:
 \begin{itemize}
     \item AC-base case: The initial Nordic 44 system in Fig.\ref{fig:N44system}.
     \item DC-segmented case: The DC-segmented system in Fig.\ref{fig:N44systemDCs} obtained with the proposed algorithm, where AC lines 5100-6500 and 3359-5101 were replaced by VSC-HVDC links. VSC-HVDC link data are in Section~\ref{sec:Appendix-N44} of the Appendix. 
 \end{itemize}

For the DC-segmented case, the operating point considered of the VSC-HVDC links is when they transmit the same active-power as the AC lines in the AC-base case, while reactive power injections equal zero at both converter stations on both links:
\begin{itemize}
    \item VSC-HVDC link 1 (VSC-1 at bus 5100, VSC-2 at bus 6500): Active power of 133 MW from VSC-1 to VSC-2. 
    \item VSC-HVDC link 2 (VSC-3 at bus 5101, VSC-4 at bus 3359): Active power of 650 MW from VSC-4 to VSC-3.
\end{itemize}

The two scenarios have been compared by means of:
%The performance of the system of the two cases will be compared by means of:
\begin{itemize}
    \item Small-signal stability analysis
    \item Non-linear time-domain simulation
\end{itemize}

\subsubsection{Small-signal analysis} \label{results_validation_SSA}
\begin{figure}[htb!]
\centering
\includegraphics[width=0.48\textwidth,trim={0.7cm 0.2cm 0.7cm 0.7cm},clip]{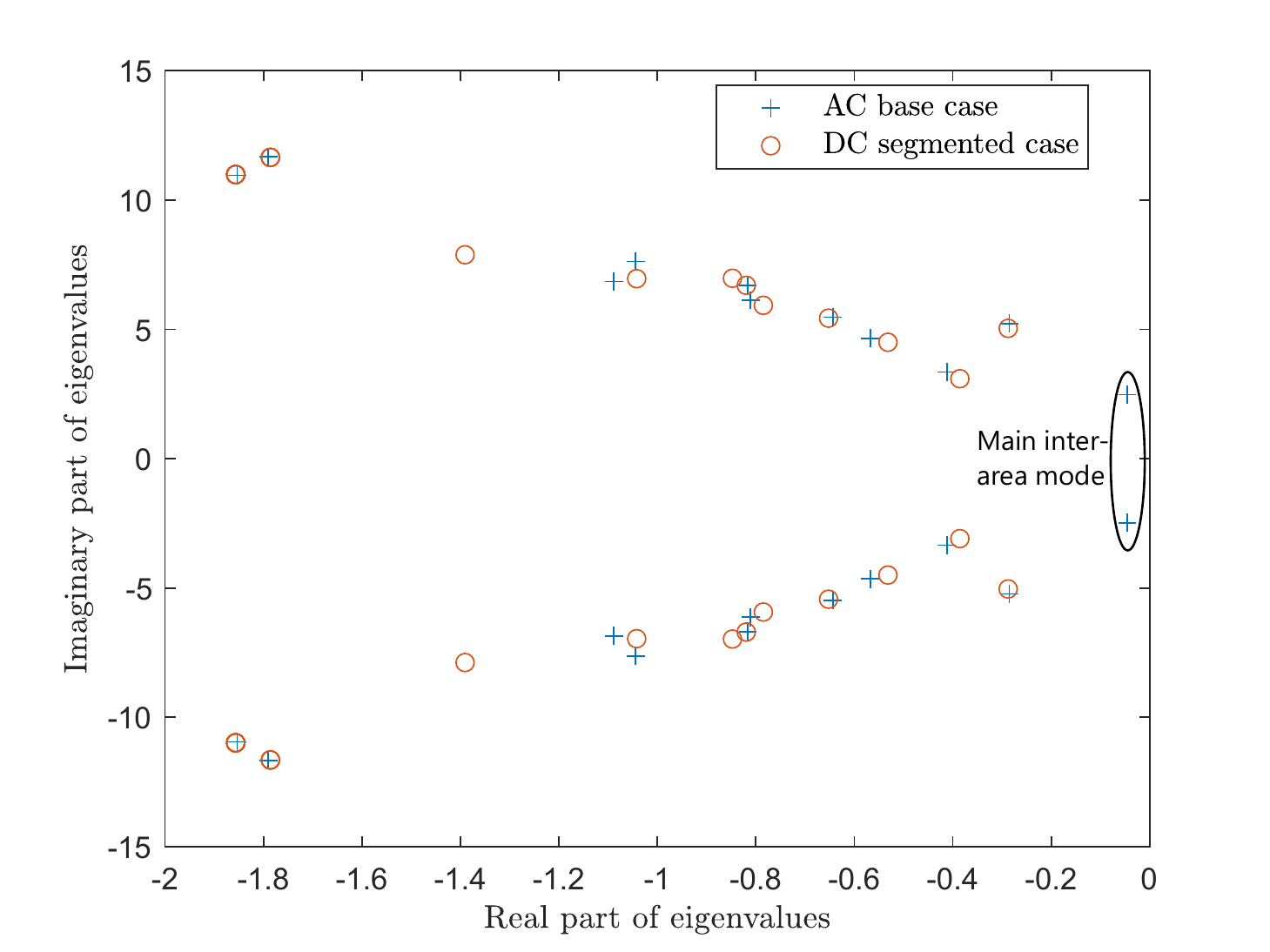}
\caption{Weakly damped electromechanical modes of the two scenarios}
\label{fig:eigvN44}
\end{figure}

Fig. \ref{fig:eigvN44} and Table \ref{tab:eigvN44} show the electromechanical modes of the system with damping ratio under 20\% obtained for the AC-base case and for the DC-segmented case. Results confirms that the critical inter-area mode (mode 1 of Table~\ref{tab:eigvN44}, with damping of 1.85\% and frequency of 0.39 Hz) has been suppressed by the DC segmentation. Meanwhile, the damping ratios of the rest of the electromechanical modes do not present significant differences between the two cases. 

\begin{table}[htb!]
%\resizebox{\columnwidth}{!}
\centering
\caption{\label{tab:eigvN44} Weakly damped electromechanical modes of the two scenarios}
\scalebox{0.75}{
\begin{tabular}{@{}c|cc|cc@{}}
\toprule
    &   \multicolumn{2}{c}{AC base case} &  \multicolumn{2}{c}{DC-segmented case} \\
N0. &   Damp (\%) & Freq (Hz)&  Damp (\%) & Freq (Hz) \\ \midrule
1	&	1.85	&	0.39	&	-	&	-	\\
2	&	5.45	&	0.83	&	5.70	&	0.80	\\
3	&	12.22	&	0.54	&	12.38	&	0.50	\\
4	&	12.11	&	0.75	&	11.73	&	0.72	\\
5	&	11.68	&	0.88	&	11.92	&	0.87	\\
6	&	13.12	&	0.98	&	13.12	&	0.95	\\
7	&	12.12	&	1.07	&	12.14	&	1.07	\\
12	&	-	&	-	&	12.06	&	1.12	\\
8	&	13.57	&	1.23	&	14.81	&	1.12	\\
9	&	15.69	&	1.10	&	17.37	&	1.27	\\
10	&	15.17	&	1.88	&	15.15	&	1.88	\\
11	&	16.69	&	1.77	&	16.66	&	1.77	\\
\bottomrule
\end{tabular}%
}
\end{table}

\FloatBarrier
\subsubsection{Non-linear time-domain simulation} \label{results_validation_TDsim}
Two faults were simulated, one in each region of the Nordic 44 system (Fig. \ref{fig:N44system} and\ref{fig:N44systemDCs}):
\begin{itemize}
    \item Fault 1 (in region 1): Three-phase-to-ground short circuit at line 6001-5402 (close to bus 6001), cleared 200 ms later by disconnecting the line. The fault occurs at $t=1$ s.
    \item Fault 2 (in region 2): Three-phase-to-ground short circuit at line 7000-7001 (close to bus 7000), cleared 200 ms later by disconnecting the three circuits of the line. The fault occurs at $t=1$ s.
\end{itemize}

\begin{figure} [htb!]
\vspace{-0.3cm}
     \centering
     \begin{subfigure}[b]{0.4\textwidth}
         \centering
         \includegraphics[width=\textwidth,trim={0.7cm 0.25cm 0.7cm 0.4cm},clip]{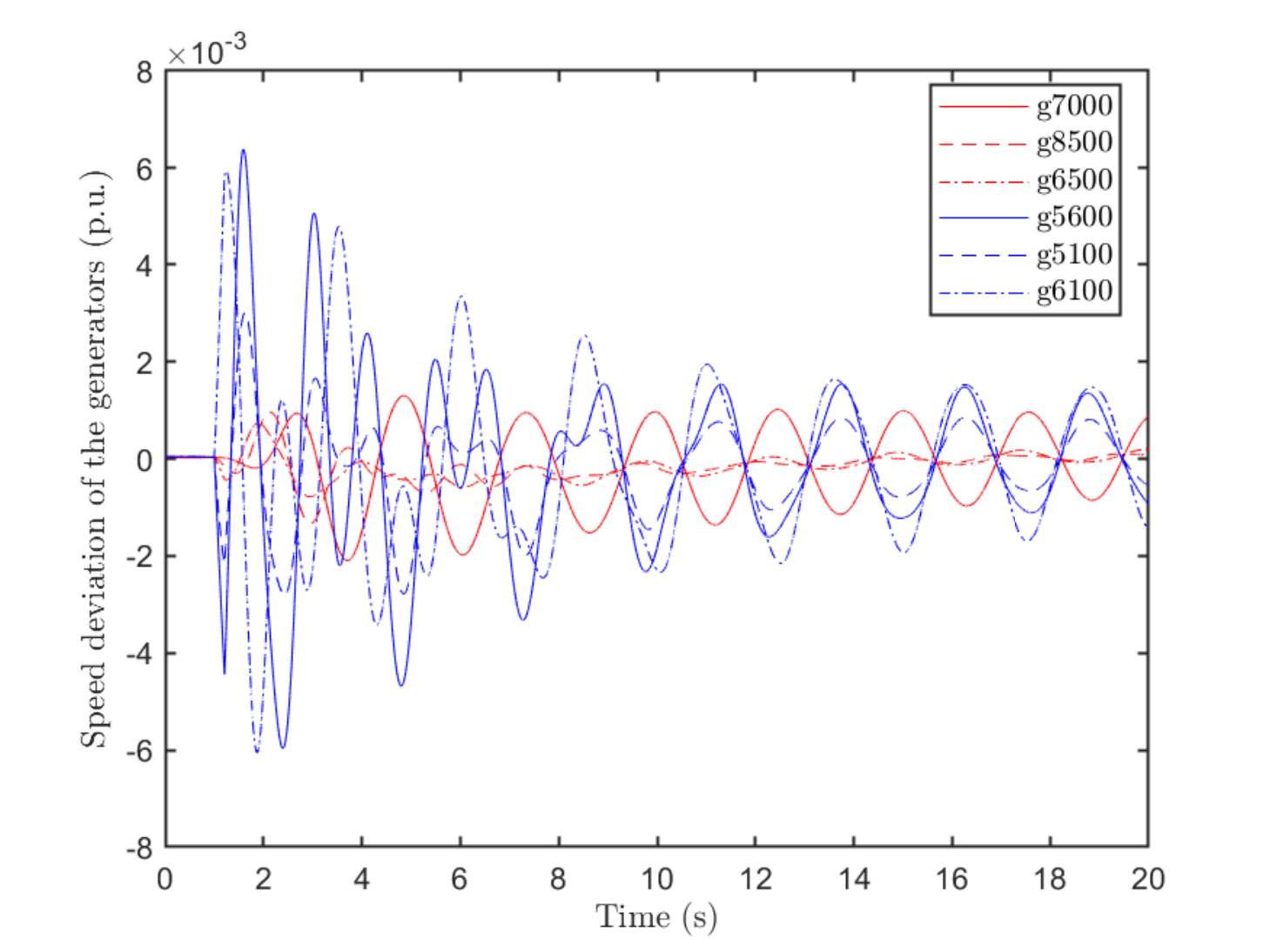}
         \caption{AC base case}
         \label{fig:fAC6001}
     \end{subfigure}
     \hfill
     \begin{subfigure}[b]{0.4\textwidth}
         \centering
         \includegraphics[width=\textwidth,trim={0.7cm 0.25cm 0.7cm 0.4cm},clip]{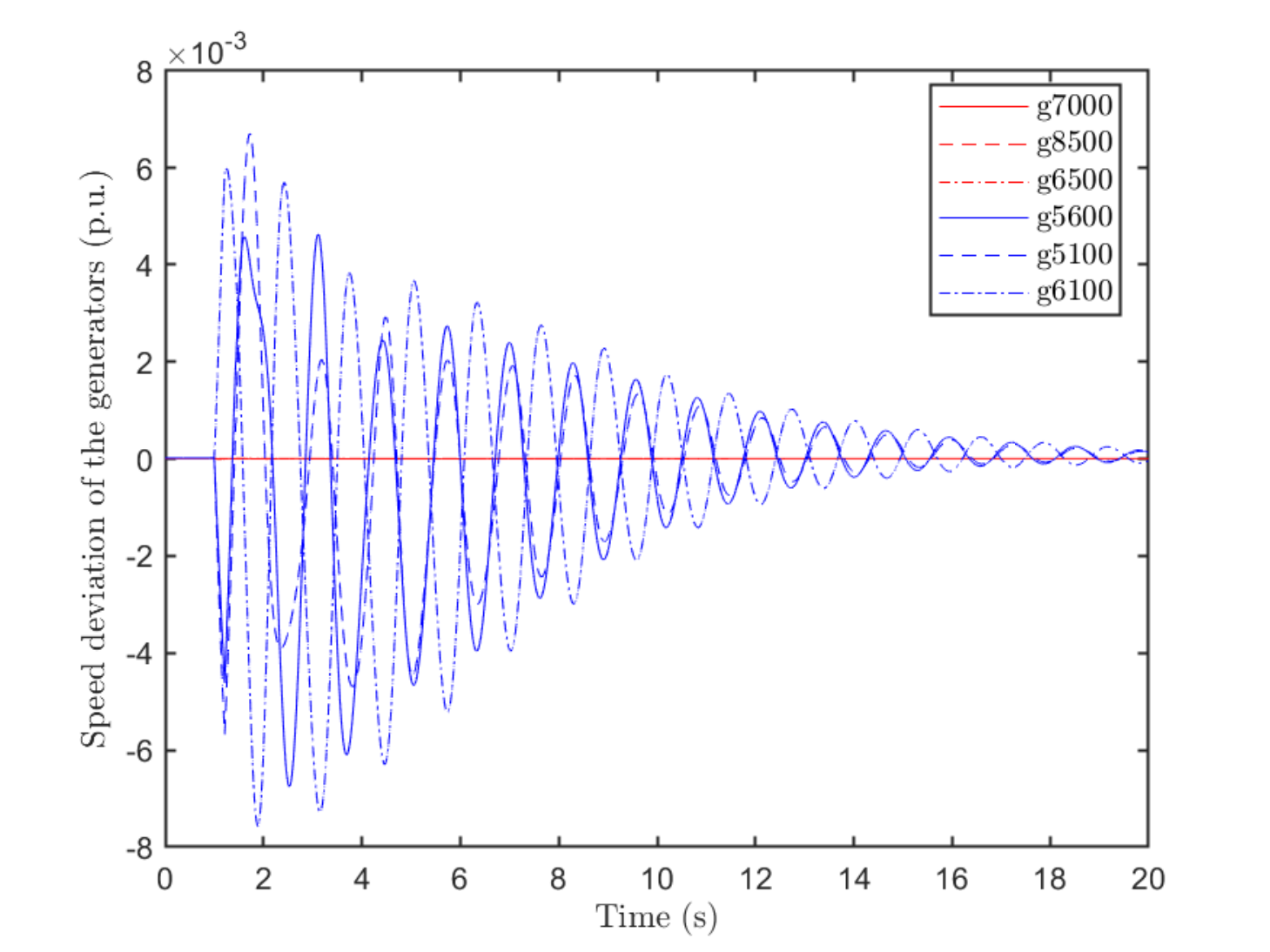}
         \caption{DC-segmented case}
         \label{fig:fDCs6001}
     \end{subfigure}
     \hfill
        \caption{Fault 1. Speed deviations of generators.}
        \label{fig:f6001}
\end{figure}

Fig.~\ref{fig:f6001} shows the speed deviations of some generators of the system when Fault 1 occurs, for the AC-base case and for the DC-segmented case. The generators have been selected to be representative of the system: three generators are in region 1 (generators 5600, 5100 and 6100 represented in blue) and the other three are in region 2 (generators 7000, 8500 and 6500 represented in red).  

In the AC base case (Fig.\ref{fig:fAC6001}), all the generators are affected by the fault and there is a poorly damped inter-area oscillation between the generators of the two regions corresponding to the targeted critical inter-area mode (mode 1 of Table \ref{tab:eigvN44}).

In the DC-segmented case (Fig.\ref{fig:fDCs6001}), generators 7000, 8500, 6500 are not affected by the fault because the VSC-HVDC links act as a firewall between regions 1 and 2. Additionally, the critical inter-area
oscillation is not present. Notice that some oscillations are still present due to the rest of electromechanical modes but they are not critical. 

\begin{figure} [htb!]
\vspace{-0.3cm}
     \centering
     \begin{subfigure}[b]{0.4\textwidth}
         \centering
         \includegraphics[width=\textwidth,trim={0.7cm 0.25cm 0.7cm 0.4cm},clip]{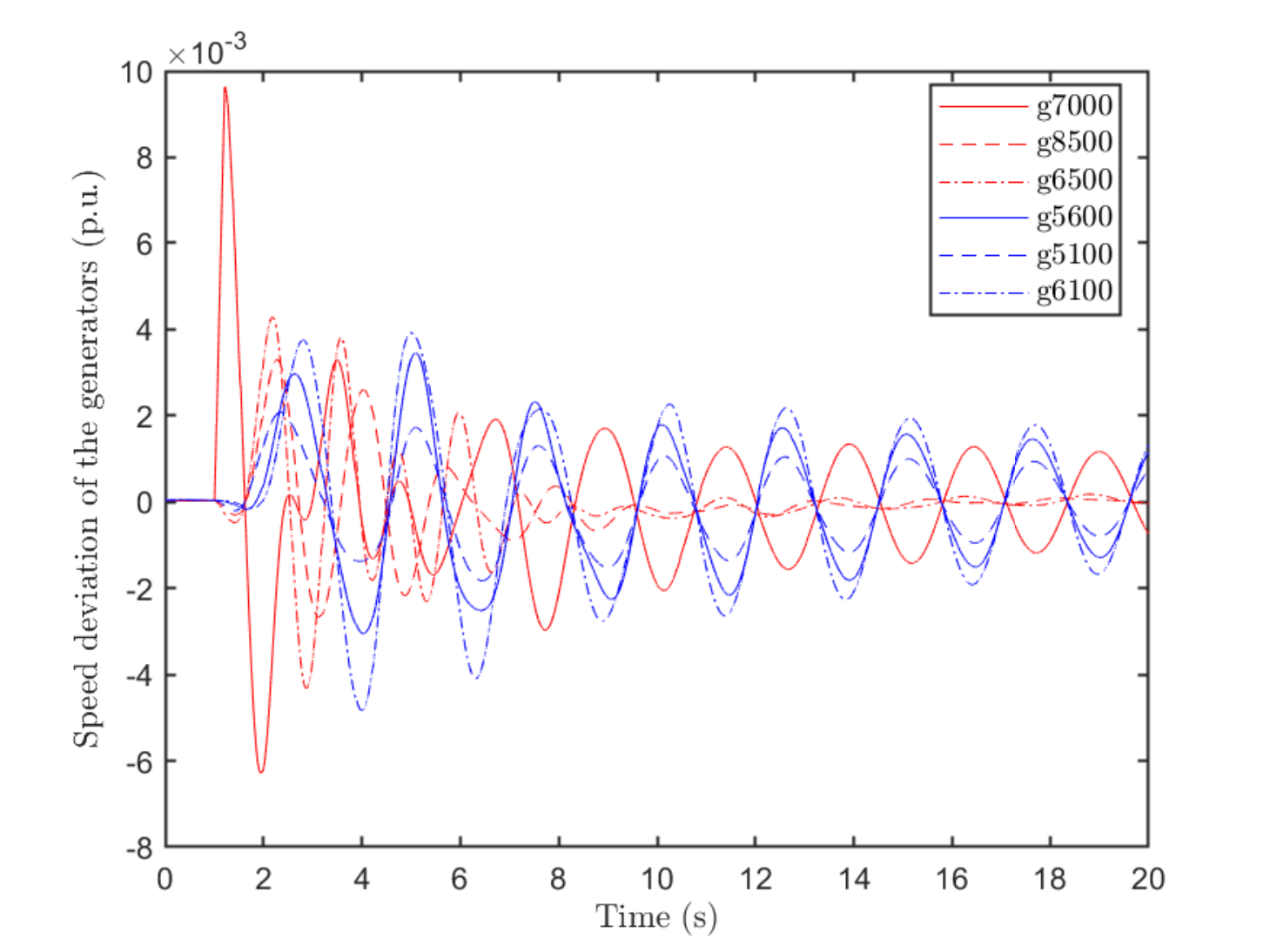}
         \caption{AC base case}
         \label{fig:fAC7000}
     \end{subfigure}
     \hfill
     \begin{subfigure}[b]{0.4\textwidth}
         \centering
         \includegraphics[width=\textwidth,trim={0.7cm 0.25cm 0.7cm 0.4cm},clip]{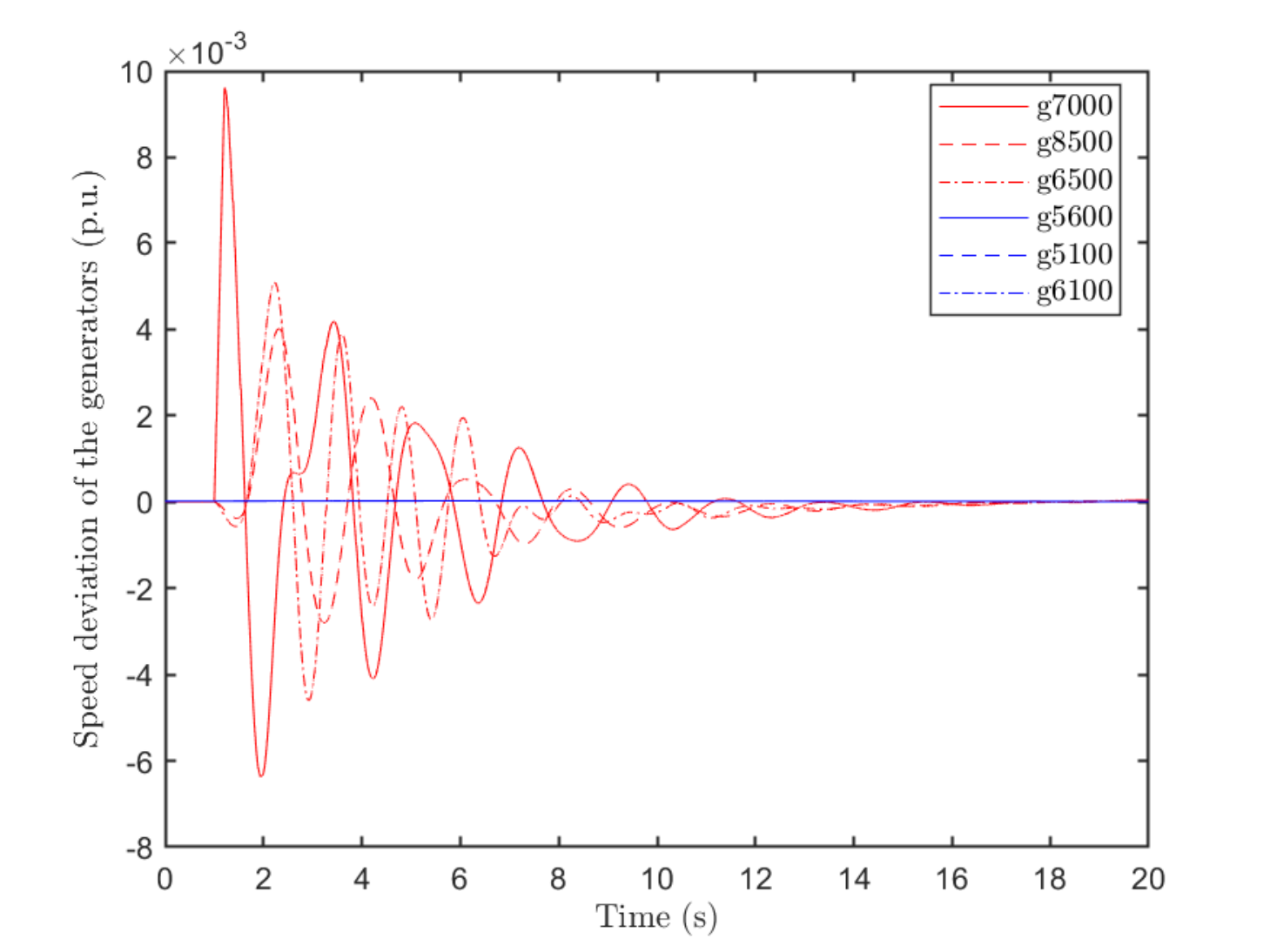}
         \caption{DC-segmented case}
         \label{fig:fDCs7000}
     \end{subfigure}
     \hfill
        \caption{Fault 2. Speed deviations of generators.}
        \label{fig:f7000}
\end{figure}

Fig.~\ref{fig:f7000} shows the results of Fault 2. The speed deviations of the same generators as before are showed. Again, in the AC base case (Fig.~\ref{fig:fAC7000}), the fault excites the speed of all generators and the critical inter-area oscillation is poorly damped. On the contrary, in the DC-segmented case (Fig. \ref{fig:fDCs7000}), the fault does not propagate from region 2 to region 1 due to the VSC-HVDC segments, and the critical inter-area oscillation is not present.

\section{Conclusions}\label{sec:conclusions}
This paper proposed an algorithm for DC segmentation of AC power systems to mitigate electromechanical oscillations, using VSC-HVDC links. The objective of the proposed algorithm is to suppress the most critical inter-area oscillation of the initial AC power system. The proposed algorithm uses information of the small-signal stability analysis of the system and the concept of inter-area oscillation paths. 

Results show that:
\begin{itemize}
    \item The proposed algorithm obtains systematically a DC-segmentation scheme of the initial AC power system.
    \item In the resulting DC-segmented scheme, the critical inter-area oscillation is suppressed, without jeopardising the damping ratio of other modes.
    \item The concept of inter-area oscillation path is a a remarkably useful tool to determine where to segment the system with DC technology, at least when tackling electromechanical oscillations.
    \item The algorithm proposed to identify the inter-area oscillation path (Step 2 of the main algorithm) obtains the path systematically and it could be used for other applications, different from DC segmentation.
\end{itemize}

% or


\appendices

\section*{Appendix}\label{sec:Appendix}

\subsection{Six-generator test system 1}\label{sec:Appendix-6g}
Each of the six generators radially connected in this system is  rated at 200~MVA and 20~kV with a nominal frequency of 50~Hz. The generator parameters in per unit on the rated MVA and kV base are: 
\begin{equation*}
\begin{matrix}
H=6.5 s, \;\; D=0 p.u., \;\; R=0.0025 p.u.,\\
T'_{d0}=8s, \;\; T''_{d0}=0.03s, \;\; T'_{q0}=0.4s, \;\; T''_{q0}=0.05s,\\
X_{d}=1.8 p.u., \;\; X_{d}=1.7 p.u., \;\; X'_{d}=0.3 p.u., \\
X'_{d}=0.55 p.u., \;\; X''_{d}=0.25 p.u., \;\; X_{L}=0.2 p.u. \\
\end{matrix} 
\end{equation*}

Each generator has an excitation system with parameters: $T_{r}=0.01s, \;\; K_{a}=200 p.u.$
Each step-up transformer has an impedance of 0 + j0,15 p.u. on a 200 MVA and 20/220 kV base and has an off-nominal ratio of 1.0.

The transmission system nominal voltage is 220 kV. The inter-area lines (lines 30-35 and 35-40) have a length of 50 km while the local lines (lines 10-20, 20-30, 10-50, and 50-60) have a length of 25 km. The parameters of the lines in p.u. per km on a 100 MVA, 220 kV base are:
\begin{equation*}
\begin{matrix}
r=0.0001 \,p.u./km, \;\; x_{L}=0.001 \,p.u./km,\\
 b_{c}=0.00175\, p.u./km\\
\end{matrix} 
\end{equation*}

The system is operating with the left area exporting 600 MW to the right area and the generating units are loaded as listed in Table~\ref{tab:PF6g}.

\begin{table}[htb!]
\centering
\caption{Initial power flow data of the six-generator system.} \label{tab:PF6g}
\scalebox{0.6}{
\begin{tabular}{@{}ccccc@{}}
\toprule
	& $P_{G,i}$ (MW)	& $Q_{G,i}$ (Mvar)	& $V_i$ (p.u.)	&  $\theta_i$ (deg)  \\ \midrule
G1	& 112.9	& 11.6	& 1	& 0.0   \\
G2	& 100	& 15.2	& 1	& -2.2  \\
G3	& 100	& 24.7	& 1	& -5.2  \\
G4	& -100	& 36.7	& 1	& -32.7 \\
G5	& -100	& 30.2	& 1	& -35.8 \\
G6	& -100	& 26.9	& 1	& -37.3 \\
\end{tabular}%
}
\end{table}

\subsection{Six-generator + one-load test system 2}\label{sec:Appendix-6g1L}
This system is like the previous one but with a 600~MW-0~MVAr load connected at bus 35.
The generating units are loaded as listed in table \ref{tab:PF6g1L}.

{\begin{table}[htb!]
\centering
\caption{Initial power flow data of the six-generator 1 load system.} \label{tab:PF6g1L}
\scalebox{0.6}{
\begin{tabular}{@{}ccccc@{}}
\toprule
	& $P_{G,i}$ (MW)	& $Q_{G,i}$ (Mvar)	& $V_i$ (p.u.)	&  $\theta_i$ (deg) \\ \midrule
G1	& 112.8	& 16.7	& 1	& 0.0  \\
G2	& 100	& 22.0	& 1	& -2.2 \\
G3	& 100	& 35.4	& 1	& -5.3 \\
G4	& 100	& 34.4	& 1	& -5.7 \\
G5	& 100	& 21.1	& 1	& -2.8 \\
G6	& 100	& 15.0	& 1	& -1.3 \\
\end{tabular}%
}
\end{table}}

\subsection{DC-segmented six-generator test system}\label{sec:Appendix-6gDCs}
The characteristics of the VSCs and the two DC lines used to segment the six-generator system are included in Table~\ref{tab:VSCdata6g}. 

{
\begin{table}[htb!]
\centering
\caption{\label{tab:VSCdata6g}Data of the HVDC links in the DC-segmented six-generator system.}
\scalebox{0.65}{
\begin{tabular}{p{5cm}p{4cm}}
\toprule
Parameters  \\					
(p.u's: converter rating) 	&	Values	\\ \midrule			
Rating VSC, DC voltage	&	500	MVA, $\pm$	320	kV	\\
Configuration	&	Symmetrical monopole	\\			
Max active (reactive) power	&	$\pm 500 MW$, ($\pm 200 MVAr$)	\\			
Max. current	&	1 p.u.	\\			
Max. DC voltage $V^{max}_{dc,i}, V^{min}_{dc,i}$	&	1.1	, 	0.9	p.u.	\\	
Current-controller time constant ($\tau$)	&	2 ms	\\			
Connection imp. ($R_{s,i} + jX_{s,i}$)	&	0.02	 + j 	0.2	p.u.	\\
Outer control gains	\\				
P prop./int.($K_{p,d1}/K_{i,d1}$)	&	0/0	\\			
V$_{dc}$ prop./int.($K_{p,d2}/K_{i,d2}$)	&	10 p.u./20 p.u./s	\\			
Q prop./int.($K_{p,q1}/K_{i,q1}$)	&	0/0	\\			
%V$_{ac}$ prop./int.($K_{p,q2}/K_{i,q2}$)	&	15 p.u./300 p.u./s	\\			
VSCs' loss coefficients	&	a = b = c = 0 p.u.	\\			
DC-bus capacitance ($C_{dc,i}$)	&	195	$\mu F$	\\		
DC-line series resistance, inductance ($R_{dc,ij}, L_{dc,ij}$)	\\				
 - line 35-40	&	2.2	$\Omega$,	77.1	mH	\\
 - line 40-50	&	1.1	$\Omega$,	38.5	mH	\\
\end{tabular}%
}
\end{table}
}

\subsection{N44 test system}\label{sec:Appendix-N44}
\subsubsection{AC base case}\label{sec:Appendix-N44_AC_base_case} 
Dynamic and static data of the Nordic 44 test system can be found under \cite{Nordic44NordpoolN44BC}. The initial power flow condition used in this paper correspond to the Nord Pool data of Tuesday November 10 at 11:38 available at the same link.

\subsubsection{DC-segmented case}\label{sec:Appendix-N44_DCs}
The DC-segmented case was obtained from the initial Nordic 44 system by replacing AC line 5100-6500 and the two parallel lines between buses 3359 and 5101 with VSC-HVDC links. The characteristics of the four VSCs and the two DC lines are included in Table \ref{tab:VSCdataN44}. VSC 5100, 6500, 5101 and 3359 are renamed VSC 1, 2, 3 and 4 in the table for simplicity.

{
\begin{table}[htb!]
\centering
\caption{\label{tab:VSCdataN44}Data of the HVDC links in the DC-segmented Nordic 44 system.}
\scalebox{0.65}{
\begin{tabular}{p{6cm}p{4cm}}
\toprule
Parameters 	\\		
(p.u's: converter rating) 	&	Values	\\ \midrule
Rating VSC (VSC1-2/ VSC3-4)	&	800/3500 MVA	\\
DC voltage (VSC1-2/ VSC3-4)	&	$\pm$ 320/535 kV	\\
Configuration	&	Symmetrical monopole	\\
Max active power (VSC1-2/ VSC3-4)	&	$\pm 800/3500 MW$	\\
Max reactive power (VSC1-2/ VSC3-4)	&	$\pm 320/1400 MVAr$	\\
Max. current	&	1 p.u.	\\
Max. DC voltage $V^{max}_{dc,i}, V^{min}_{dc,i}$	&	1.1, 0.9 p.u.	\\
Current-controller time constant ($\tau$)	&	2 ms	\\
Connection imp. ($R_{s,i} + jX_{s,i}$)	&	0.004 + j 0.2 p.u.	\\
Outer control gains	&	\\	
P prop./int.($K_{p,d1}/K_{i,d1}$)	&	0/0	\\
V$_{dc}$ prop./int.($K_{p,d2}/K_{i,d2}$)	&	10 p.u./20 p.u./s	\\
Q prop./int.($K_{p,q1}/K_{i,q1}$)	&	0/0	\\
%V$_{ac}$ prop./int.($K_{p,q2}/K_{i,q2}$)	&	15 p.u./300 p.u./s	\\
VSCs' loss coefficients	&	a = b = c = 0 p.u.	\\
DC-bus capacitance ($C_{dc,i}$) (VSC1-2/ VSC3-4)	&	195/305 $\mu F$	\\
DC-line series resistance, inductance ($R_{dc,ij}, L_{dc,ij}$)	&	\\	
 - line 5100-6500	&	7.2 $\Omega$, 258 mH	\\
 - line 5101-3359	&	1.6 $\Omega$, 67 mH	\\
 \\
\end{tabular}%
}
\end{table}
}
VSCs 6500 and 3359 were set in slave mode (control of DC voltage and reactive power injection) while VSCs 5100 and 5101 were set in master mode (control of active-and reactive-power injections).

\subsection{Algorithms to identify the dominant inter-area oscillation path}\label{sec:Appendix-inter_area_oscillation_path_comparison}
The main characteristics of the algorithm to identify the inter-area oscillation path proposed in \cite{chompoobutrgoolIdentificationPowerSystem2013} can be summarised as follows:
\begin{itemize}
    \item It uses small-signal stability analysis and bus-angle, bus-voltage and branch-current observability factors. 
    %\item The inter-area oscillation path is determined by sorting the magnitude of the branch-current observability factors ($|\phi_{I_{ij}}|$) in descending order.
    \item Branches with high values of the current observability factors $|\phi_{I_{ij}}|$ are included in the inter-area oscillation path.
    \item Then, the user analyses the single-line-diagram of the system, the branches initially included in the path and the information provided by the observability factors, and makes a decision on the inter-area oscillation path.
\end{itemize}

The proposed algorithm to identify the inter-area oscillation path (Step 2 of the main algorithm, see Section \ref{sec:Step2bis}), can be summarised as follows:
\begin{itemize}
    \item It uses small-signal stability analysis, generator-speed mode shapes, bus-frequency and branch-current observability factors. 
    \item The algorithm proposed in this work is incremental: starting from the first edge of the path (E1), it gradually finds all the branches and buses of the path, one by one, until the second edge of the path (E2) is reached.
    \item The path is found using not only branch-current observability factors, but also bus-frequency observability factors. 
    \item The inter-area oscillation path is fully determined by the algorithm, without additional analysis. 
\end{itemize}

\section*{Acknowledgment}
The authors would like to thank Prof. Luis Rouco for sharing SSST tool and interesting discussions on electromechanical oscillations, Prof. Luigi Vanfretti for developing OpenIPSL and interesting discussions on electromechanical oscillation paths and Dr. Lukas Sigrist for interesting discussion on small signal stability.

The work of Mr. Mathieu Robin is within a collaboration of SuperGrid institute in the doctoral programme of Comillas Pontifical University.


% \bibliographystyle{IEEEtran}
% \bibliography{articleAlgo.bib}


\begin{IEEEbiography}[{\includegraphics[width=1in,height=1.25in,clip,keepaspectratio]{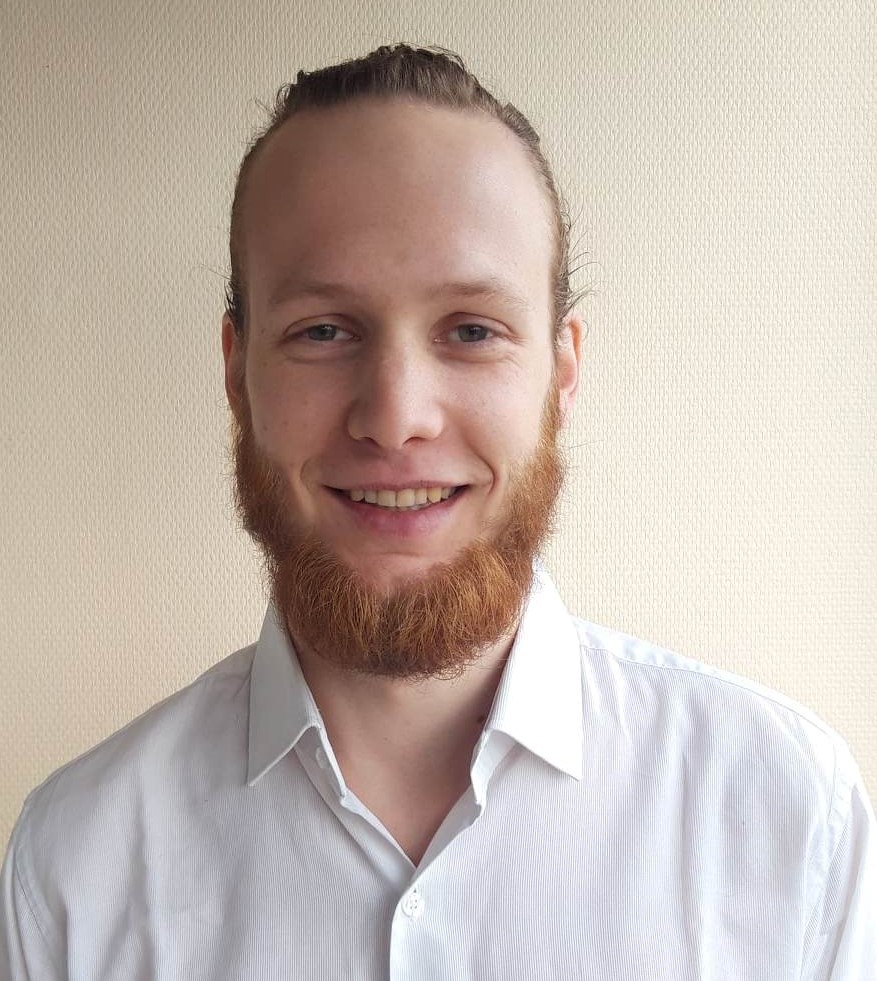}}]{Mathieu Robin}
obtained the M.Eng degree from Ecole Centrale Lyon, Ecully, France  and  the  M.Sc  degree  in Electrical Engineering from jointly Ecole Centrale Lyon and Claude Bernard University in Lyon, France, both in 2019. He is actually pursuing his Ph.D. degree at Universidad Pontificia Comillas de Madrid (UPCO), Madrid, Spain and is working as a research engineer at SuperGrid Institute, Villeurbanne, France. 

His research interests are stability, modelling and control of power systems, particularly VSC-HVDC systems and DC segmentation.
\end{IEEEbiography}

\begin{IEEEbiography}[{\includegraphics[width=1in,height=1.25in,clip,keepaspectratio]{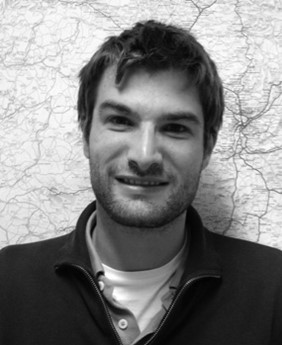}}]{Javier Renedo}
(Senior Member, IEEE) received his M.Sc. degree in electrical engineering from the Universidad Pontificia Comillas de Madrid (UPCO), Madrid, Spain, in 2010, the M.Sc. degree in mathematical engineering from the Universidad Carlos III de Madrid, Madrid, Spain, in 2013, and the Ph.D. degree in modelling of engineering systems from UPCO, in 2018. His research interests include power system stability, VSC-HVDC systems, and power systems with large amounts of renewable resources.
\end{IEEEbiography}
\begin{IEEEbiography}
[{\includegraphics[width=1in,height=1.25in,clip,keepaspectratio]{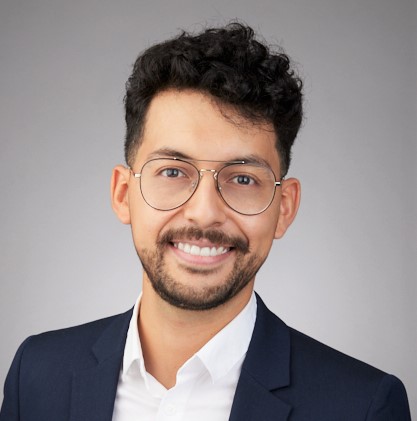}}]{Juan Carlos Gonzalez Torres}
received both the Electromechanical Engineering degree from Universidad Autonoma de San Luis Potosi and the Master of Engineering from Ecole Centrale de Lille in 2014. He graduated with a Master of Science degree from ParisTech  (The Paris Institute of Technology)  in 2015. In 2019, he obtained the Ph.D. in Automatic Control from Paris-Saclay University.  He has been working as a research engineer at Supergrid Institute since 2016,  where  he is part of the Architecture \& Systems program.  His main research interests include modeling and control of power systems, HVDC transmission systems and integration of renewable energies via power electronics-based devices.
\end{IEEEbiography}

\begin{IEEEbiography}[{\includegraphics[width=1in,height=1.25in,clip,keepaspectratio]{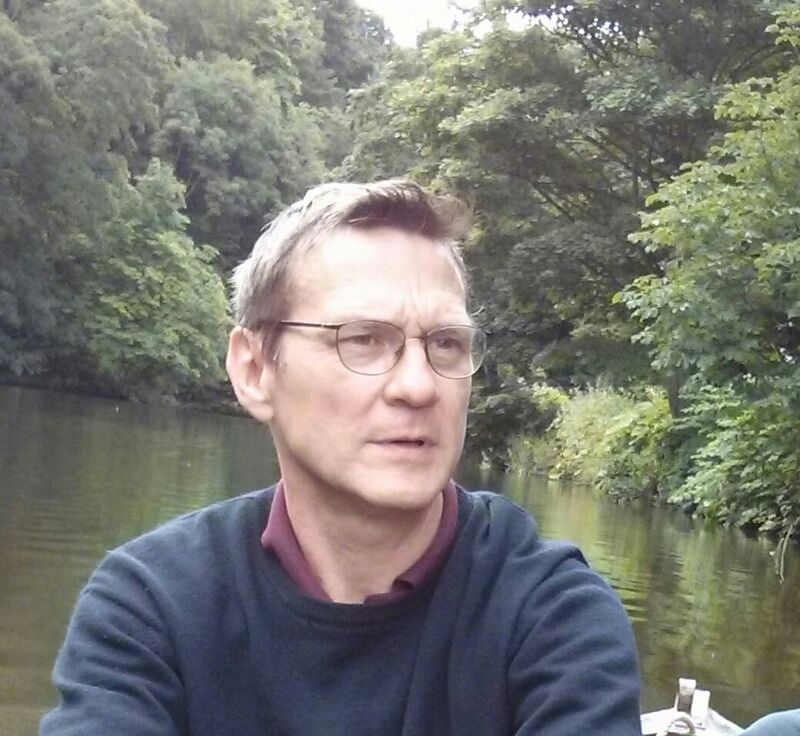}}]{Aurelio Garcia Cerrada}
 (Senior Member, IEEE from 2015) M.Sc. (1986) from the Universidad Politécnica de Madrid, Spain, and Ph.D. (1991) from the University of Birmingham, U.K. He is a Professor in the Electronics, Control Engineering and Communications Department and a member of the Institute for Research in Technology (IIT) at the Universidad Pontiﬁcia Comillas de Madrid. His research focuses on power electronics and its applications to electric energy systems.
\end{IEEEbiography}

\begin{IEEEbiography}
[{\includegraphics[width=1in,height=1.25in,clip,keepaspectratio]{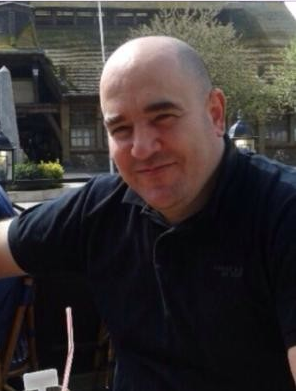}}]{Abdelkrim Benchaib}
received his Ph.D. from the Automatic Systems Laboratory, Picardie University, France, in December 1998 and in 2014 his HDR (French postdoctoral degree allowing its holder to supervise PhD students) from Paris-Orsay University.  He joined Alstom in July 2000 where he has been working as a power quality and smart grid project leader and thereafter with GE Grid solutions. Currently, he is seconded by GE to work with the SuperGrid institute where he is a Sub-program group leader for real-time strategies of super grids AC/DC control and dispatch. His expertise and research interests include automatic control, AC and DC power systems and power electronics. Dr. Benchaib is associate Professor at the Cnam (Conservatoire National des Arts et Metiers) teaching wind energy and power network. He has been General Chairman of the EPE ECCE Europe Conference for its 2020 edition (EPE conference is one of the bigest events in the world for Power Electronics with up to 1000 participants) . Abdelkrim Benchaib is the secretary of the EPE association.
\end{IEEEbiography}

\begin{IEEEbiography}[{\includegraphics[width=1in,height=1.25in,clip,keepaspectratio]{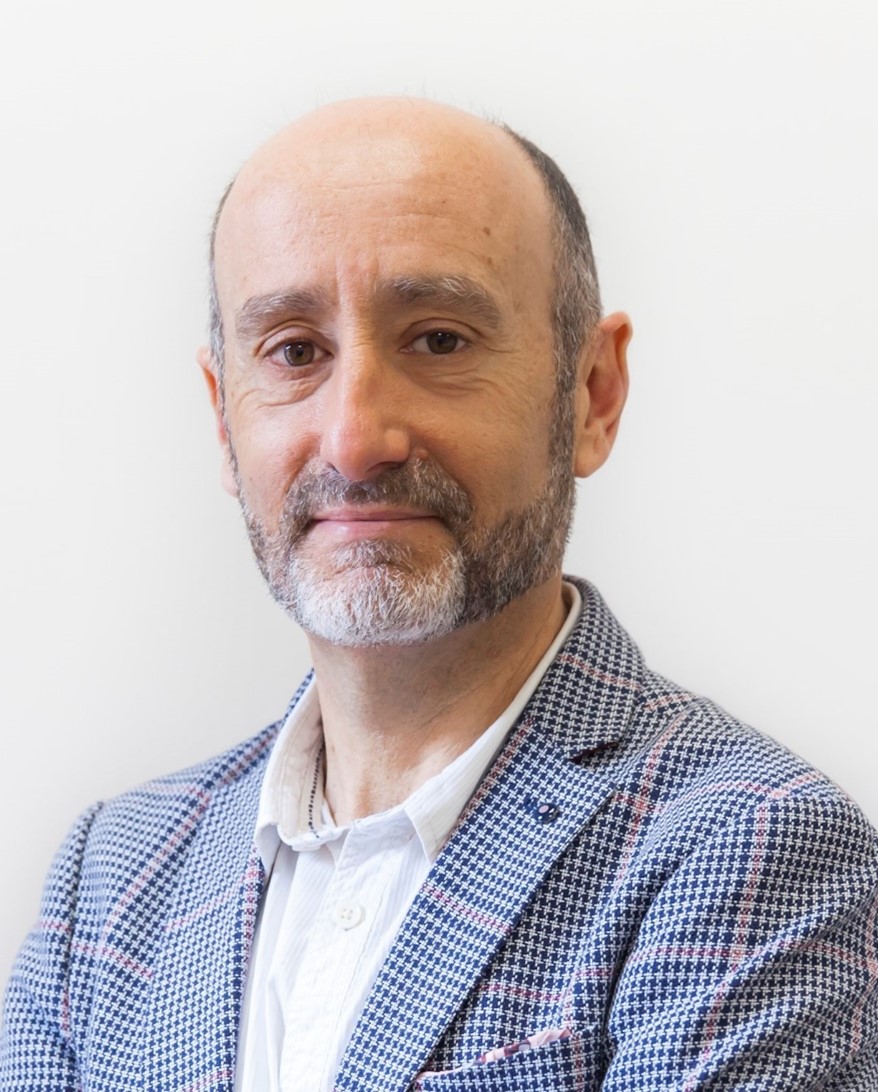}}]{Pablo Garcia Gonzalez}
has a M.Sc. in electrical engineering (1992) and a PhD (2000) from the Universidad Pontificia Comillas, Madrid, Spain. He is Full Professors of the ICAI School of Engineering. His teaching and research activities focuses on applications of power electronics, control systems and integration of distributed energy resources in power systems.
\end{IEEEbiography}


%\vfill


\end{document}